\documentclass[11pt]{article}

\title{Numerical approach to the modular operator for fermionic systems}
\author{%
	Henning Bostelmann\footnote{Fachbereich Ingenieur- und Naturwissenschaften, Hochschule Merseburg, Eberhard-Leibnitz-Stra{\ss}e 2, 06217 Merseburg, Germany; e-mail: \url{henning.bostelmann@hs-merseburg.de}}, 
	Daniela Cadamuro\footnote{Institut f\"ur theoretische Physik, Universit\"at Leipzig, Br\"uderstra\ss e 16, 04103 Leipzig, Germany; e-mail: \url{cadamuro@itp.uni-leipzig.de}}, 
	Christoph Minz\footnote{Institut f\"ur theoretische Physik, Universit\"at Leipzig, Br\"uderstra\ss e 16, 04103 Leipzig, Germany; e-mail: \url{christoph.minz@itp.uni-leipzig.de}}}
\date{}

\usepackage{paper}

\addto\extrasenglish{%
}

\usepackage{csquotes}  
\usepackage[%
backend=biber,
style=alphabetic,  
citestyle=alphabetic,  
sorting=nyt,
maxnames=10,  
giveninits=true,  
eprint=true,  
doi=false,  
url=false  
]{biblatex}
\graphicspath{{figures/}} 

\begin{document}

\maketitle

\begin{abstract}
We numerically approximate the Tomita-Takesaki modular operator for local subalgebras of the 1+1-dimensional massive Majorana field.
Our method works at the one-particle level with a discretisation of time-0 data in position space. 
The local subspaces we consider are associated with one double cone and with the disjoint union of two double cones. In order to avoid boundary effects, we primarily choose the overall spacetime to be a cylinder; different choices of boundary conditions (antiperiodic and periodic) are considered. 
We compare our numerical results to known analytic expressions in the massless case. It turns out that the modular operator has a non-trivial dependence on the mass. In the case of two double cones, the modular generator does not only have ''local'' contributions (supported on the diagonal in configuration space) but also ''bilocal'' terms (connecting the two double cone regions); we find the latter to be less prominent at higher masses, in line with expectations.
\end{abstract}

\section{Introduction}

Tomita-Takesaki modular theory can be regarded as one of the most significant developments in the field of operator algebras since 1970, in particular with its application in Connes' classification of von Neumann factors \cite{Connes:1973}. It also gained importance in quantum physics, where the modular operator $\Delta$ replaces the usual trace matrix $\rho$ in the description of states on generic von Neumann algebras, in particular on those of type III. This has led to a wealth of applications in thermodynamics  \cite{HaagHugenholtzWinnink:1967} and quantum field theory, both conceptual --- see \cite{Borchers:2000} for a review --- and constructive \cite{Lechner:2008}. The modular operator also plays a role in describing information-theoretic quantities such as the (relative) entropy of states \cite{Araki:1976}, which has recently led to renewed interest in the context of relativistic quantum information \cite{AriasCasiniHuertaPontello:2018,  CasiniGrilloPontello:2019,CiolliLongoRuzzi:2020}.

Despite its importance, it is often less clear how the operator $\Delta$ can be explicitly described in examples. In the thermodynamic setting \cite{HaagHugenholtzWinnink:1967}, $\log \Delta$ is the generator of time evolutions (up to a numerical factor), whereas for the algebra of a wedge region in quantum field theory it is connected to the boost symmetry (Bisognano--Wichmann theorem \cite{BisognanoWichmann:1976}), a relation that can be extended somewhat in systems with conformal symmetry \cite{BrunettiGuidoLongo:1993}. Beyond these results, even in free (i.e.~linear) quantum field theories --- to which we will restrict in this paper ---, little is known in general. 
For \emph{massless} free theories, which have an additional symmetry, results are available for double cone \cite{HislopLongo:1982} and light cone \cite{Buchholz:1978} regions, as well as, in low dimensions, for the disjoint union of several double cones \cite{CasiniHuerta:2009,LongoMartinettiRehren:2010,RehrenTedesco:2013,KlichVamanWong:2017}; but the structure of $\Delta$ in massive free theories presents open challenges.

Our goal in this publication is to exhibit details of the modular generator $\log \Delta$ in free fermionic quantum field theories using numerical approximation, extending our previous results in the bosonic case  \cite{BostelmannCadamuroMinz:2023}. This is simplified by the fact that the modular generator of the CAR algebras is second quantized, that is, determined by a corresponding Tomita-Takesaki operator at one-particle level \cite{Foit:1983}. As we will recall in \autoref{sec:ModularTheory}, this single-particle modular generator is given as a block operator
\begin{align}
        - i \log \Delta
	&= \begin{pmatrix}
			0 & M_- \\
			- M_+ & 0
		\end{pmatrix}
    \eqend{,}
\end{align}
with the operators
\begin{align}
\label{eq:Mpm}
		M_{\pm}
	&= 2 A^{\pm\frac{1}{4}}
		\artanh\left(
		A^{+\frac{1}{4}} \chi A^{-\frac{1}{4}}
        + A^{-\frac{1}{4}} \chi A^{+\frac{1}{4}}
        - 1
        \right)
		A^{\pm\frac{1}{4}}
	\eqend{,}
\end{align}
where the operator $A$ is determined by the field equation, and $\chi$ multiplies the initial data of the field equation (in configuration space) with the characteristic function of the region in question. 
It is these operators $M_{\pm}$ that we numerically approximate, namely by discretizing their constituents $A^{\pm\frac{1}{4}}$ and $\chi$ into finite-dimensional matrices. Unlike in the bosonic case \cite{BostelmannCadamuroMinz:2023}, the argument of the $\artanh$ is a \emph{bounded} operator, but with its spectral values clustering at $\pm 1$. This requires a rather different approach to the numerical problem that we will detail in \autoref{sec:NumericalApproach}.

Specifically, we consider the following setting which we take to be indicative of the general situation, while still accessible with moderate computational power: Our model is the free Majorana fermion of mass $m \geq 0$ in $(1 + 1)$-dimensional Minkowski spacetime and on a $(1 + 1)$-dimensional flat cylinder spacetime; the subregions considered are 
\begin{enumerate}[label=(\alph*), noitemsep, topsep=-1ex]
\item the right wedge as a subregion of Minkowski space, 
\item a double cone as a subregion of the cylinder and Minkowski space, and 
\item two double cones as a subregion of the cylinder. 
\end{enumerate}(The cylinder spacetime is chosen over Minkowski spacetime to avoid numerical artefacts arising from a large-distance spatial cutoff in the discretization.)

While case (a) has a known analytic result for all masses (Bisognano-Wichmann) and can hence serve as a test case, (b) and (c) are known only in the massless case, and our goal is to study the mass dependence of $M_\pm$, also in view of previous conjectures in the bosonic case that certain components of $\log \Delta$ would be mass-independent. 

A particular aspect of the massless case is that in (b), the operator kernel of $M_-$ in configuration space is supported only on the diagonal, a feature often called ''locality'' (although it should perhaps loosely be seen as a weaker form of the principle of geometric modular action \cite{BuchholzDreyerFlorigSummers:2000}), whereas in (c), the support is larger, with a ''bilocal'' term connecting the two double cones. We investigate whether this phenomenon persist when the mass is nonzero.

Further, on the cylinder spacetime, the massless system may have a ``zero mode'' (an ambiguity of the ground state) which is partially dependent on the boundary conditions chosen in spacelike direction of the cylinder.
The massive case, however, does not, which makes the massless limit of our system particularly interesting.
We compare the behaviour of our numerical data in the limit $m \to 0$ to the known analytic results at $m = 0$.

The paper is organised as follows. First, in \autoref{sec:ModularTheory}, we recall the expression of the modular generator for free Fermions, both in an abstract context and in the example of Majorana fields in $1 + 1$ spacetime dimensions. In \autoref{sec:NumericalApproach} we explain our numerical approach to evaluate $M_-$ given by \eqref{eq:Mpm}. 
In \autoref{sec:NumericalResults} we present our numerical results for one and for two double cones, and we compare this approximation to existing analytic results for the two-dimensional massless Majorana field. We conclude with a discussion and outlook in \autoref{sec:Conclusion}. Three appendices contains some details of the computation, in particular regarding relations to the existing literature. The computer code used for producing the results is provided as supplemental material to this paper.

\section{Tomita-Takesaki modular theory for free fermionic systems}
\label{sec:ModularTheory}

In this section, we briefly describe the fermionic systems in which we intend to approximate the modular operator. We restrict our attention to the structures at single-particle level, since the modular data of the CAR algebras acting on the vacuum Fock space can then be obtained directly by second quantization \cite{Foit:1983}.
We first present the generic setting (\autoref{ssec:SingleParticleStructures}) before detailing the systems in $1 + 1$ spacetime dimensions that we will consider in our computations (\autoref{ssec:FermionicSystemsD2.Minkowski}, \autoref{ssec:FermionicSystemsD2.Cylinder}).

\subsection{Single-particle structures}
\label{ssec:SingleParticleStructures}

The single-particle space of a free fermionic theory can abstractly be described as follows, in analogy to the bosonic case \cite{BostelmannCadamuroMinz:2023}. 

\begin{definition}
A \emph{fermionic one-particle structure} $( \mathcal{H}_{\mathrm{r}}, \mathcal{L}_{\mathrm{r}}, A^{1/4} )$ is a separable real Hilbert space $\mathcal{H}_{\mathrm{r}}$, a closed subspace $\mathcal{L}_{\mathrm{r}} \subset \mathcal{H}_{\mathrm{r}}$, and an orthogonal transformation $A^{1/4}$ on $\mathcal{H}_{\mathrm{r}}$ such that:
\begin{align}
\label{eq:OneParticleStructure.NonlocalityCondition}
      A^{\frac{1}{4}} \mathcal{L}_{\mathrm{r}}
    \cap A^{-\frac{1}{4}} \mathcal{L}_{\mathrm{r}}
  &= \{ 0 \}
  = A^{\frac{1}{4}} \mathcal{L}_{\mathrm{r}}^\perp
    \cap A^{-\frac{1}{4}} \mathcal{L}_{\mathrm{r}}^\perp
  \eqend{.}
\end{align}
\end{definition}
We call the orthogonal transformation $A^{1/4}$, rather than $A$, to keep the formal analogy to the bosonic case. We will choose a skew-symmetric operator $S$ such that
\begin{align}
\label{eq:Fermionic.Unitary.SkewSymmetricGenerator}
    A^{\pm\frac{1}{4}}
  &= \exp\left( \pm \frac{S}{4} \right)
  \eqend{.}
\end{align}
The real-orthogonal projector onto $\mathcal{L}_{\mathrm{r}}$ will be denoted $\chi$ in the following.

Given a fermionic one-particle structure, the complex one-particle Hilbert space is defined as $\mathcal{H} \coloneq \mathcal{H}_{\mathrm{r}} \oplus \mathcal{H}_{\mathrm{r}}$ and given the complex structure
\begin{align}
\label{eq:Fermionic.ComplexStructure}
        i_A
    &\coloneq \begin{pmatrix}
            0 & A^{-\frac{1}{2}} \\
            - A^{\frac{1}{2}} & 0
        \end{pmatrix}
\end{align}
and the complex inner product (for all $f = f_+ \oplus f_-, g = g_+ \oplus g_- \in \mathcal{H}$)
\begin{align}
\label{eq:Fermionic.InnerProduct}
        \innerProd[\mathcal{H}]{f}{g}
    &\coloneq \innerProd[+ \oplus -]{f}{g} - \i \innerProd[+ \oplus -]{f}{i_A g}
    \eqend{,}
\end{align}
where $\innerProd[+ \oplus -]{\cdot}{\cdot}$ is the direct sum of real inner products on $\mathcal{H}_{\mathrm{r}} \oplus \mathcal{H}_{\mathrm{r}}$. 
The real-orthogonal projector onto the real subspace $\mathcal{L} \coloneq \mathcal{L}_{\mathrm{r}} \oplus \mathcal{L}_{\mathrm{r}}\subset\mathcal{H}$ is then given by $E = \chi \oplus \chi$.

The ``nonlocality'' condition~\eqref{eq:OneParticleStructure.NonlocalityCondition} implies that $\mathcal{L}\subset\mathcal{H}$ is standard, i.e.\@, $\mathcal{L} \cap i_A \mathcal{L} = \{0\}$ and $\mathcal{L} + i_A \mathcal{L}$ is dense in $\mathcal{H}$. 
Thus we can define the one-particle Tomita operator $T$ as the closure of the map 
\begin{equation}
\label{eq:Fermionic.TomitaOperator}
        \mathcal{L} + i_A \mathcal{L}
    \to \mathcal{L} - i_A \mathcal{L}
    \eqend{,}
    \qquad
        h + i_A k
    \mapsto h - i_A k
    \eqend{;}
\end{equation}
its polar decomposition is written as $J \Delta^{1/2}$, where the modular operator $\Delta$ is the object of our interest. We recall the known relation between $\Delta$ and the orthogonal projection $E$ \cite{FiglioliniGuido:1994,Longo:2022},
\begin{align}
\label{eq:Fermionic.OrthogonalProjection.ModularOperator}
		  E
	&= ( 1 + T ) ( 1 + \Delta )^{-1}
	\eqend{.}
\end{align}
Using that $T$ is complex-antilinear, one finds that
\begin{align}
\label{eq:Fermionic.OrthogonalProjection.ModularOperator.Combined}
		1
		- E
		+ i_{A} E i_{A}
	&= ( \Delta - 1 ) ( \Delta + 1 )^{-1}
	= \tanh\left( \frac{1}{2} \log \Delta \right)
	\eqend{,}
\end{align}
which can be solved for the modular generator 
\begin{align}
\label{eq:Fermionic.ModularGenerator}
		\log \Delta
	&= - 2 \artanh( E - i_{A} E i_{A} - 1 )
	\eqend{.}
\end{align}
Expressing this in terms of $A$ and $\chi$, the modular generator takes the block form
\begin{align}
\label{eq:Fermionic.ModularGenerator.BlockForm}
        - i_A \log \Delta
	&= \begin{pmatrix}
			0 & M_- \\
			- M_+ & 0
		\end{pmatrix}
\end{align}
with 
\begin{align}
\label{eq:Fermionic.ModularOperator.Blocks}
		M_{\pm}
	&= 2 A^{\pm\frac{1}{4}}
		\artanh( B )
		A^{\pm\frac{1}{4}}
  \eqend{,}
\end{align}
where 
\begin{align}
\label{eq:Fermionic.ModularOperator.Blocks.artanhArg}
		B
  &= A^{+\frac{1}{4}} \chi A^{-\frac{1}{4}}
    + A^{-\frac{1}{4}} \chi A^{+\frac{1}{4}}
    - 1
\end{align}
is a bounded symmetric operator on $\mathcal{H}_{\mathrm{r}}$ with spectrum contained in $( -1, 1 )$.
As the two blocks $M_\pm$ are related by an orthogonal transformation, see \eqref{eq:Fermionic.ModularOperator.Blocks}, we will analyze the block $M_-$ only.

Note that these expressions for $\log\Delta$ are formally very similar to their analogues in the bosonic case \cite{BostelmannCadamuroMinz:2023}. By introducing a single-particle grading operator, it would certainly be possible to treat both bosonic and fermionic cases on equal footing; since however the numerical approximation methods for the orthogonal operator $A$ in the fermionic case (\autoref{sec:NumericalApproach}) will differ significantly from its self-adjoint bosonic counterpart \cite{BostelmannCadamuroMinz:2023}, a unified approach would be of limited use in the present context.

\subsection{Majorana fermions on Minkowski space}
\label{ssec:FermionicSystemsD2.Minkowski}

As an explicit example, we consider free Majorana fermions on a $(1 + 1)$-dimensional Minkowski spacetime, which we summarize here and, in \autoref{appx:Majorana.StandardFormalism}, compare with other conventions.
We choose $\mathcal{H}_{\mathrm{r}} = \Lp2( \Reals, \Reals )$, the initial data of one spinor component  ``in configuration space'', with a subspace $\mathcal{L}_{\mathrm{r}} = \{\psi \in \mathcal{H}_{\mathrm{r}} \mid \supp \psi \subset \mathcal{B}\}$, the wave functions with initial data supported in $\mathcal{B}$, where $\mathcal{B}$ is some interval or union of intervals. $\chi$ multiplies with the characteristic function of $\mathcal{B}$. The operator $A$, or $A^{\pm 1/2}$, is best described ``in momentum space'' (acting on Fourier transforms of functions $\psi\in \mathcal{H}_{\mathrm{r}}$), where it is given as a multiplication operator by
\begin{align}
\label{eq:Fermionic.Unitary.MomentumSpace}
    \hat{A}^{\pm\frac{1}{2}}(p)
  &= \frac{m \mp \i p}{\sqrt{p^2 + m^2}}
  = \exp\left( \mp \i \arctan \frac{p}{m} \right)
  \eqend{.}
\end{align}
One verifies condition \eqref{eq:OneParticleStructure.NonlocalityCondition} by means of anti-locality of the operator $(-\partial_x^2+m^2)^{-1/2}$, see \cite{SegalGoodman:1965}, with the numerator of \eqref{eq:Fermionic.Unitary.MomentumSpace} being local.
The generator $S$ multiplies ``in momentum space'' by the function $\hat S(p) = -2 \i \arctan(p / m)$;
writing it in terms of its kernel $S( x, y )$ ``in configuration space'' (acting on $\psi\in\mathcal{H}_{\mathrm{r}}$ by convolution), we have in the sense of distributions, 
\begin{subequations}
\eqseqlabel{eq:Fermionic.SkewSymmetricGenerator}
\begin{align}
\label{eq:Fermionic.SkewSymmetricGenerator.Integral}
    S(x, y)
  &= - \frac{\i}{\pi}
    \lim_{\varepsilon \to 0^+} \int_{-\infty}^{\infty}
      \arctan\left( \frac{p}{m} \right)
      \e^{\i p (x - y) - \varepsilon \abs{p}}
    \id{p}
\\\label{eq:Fermionic.SkewSymmetricGenerator.PV}
  &= \pvalue \frac{1}{x - y} \e^{- m \lvert x - y \rvert}
\end{align}
\end{subequations}
where the integral follows from~\cite[Eq.~(3), p.~87]{Bateman:1954}.

\subsection{Majorana fermions on a cylinder}
\label{ssec:FermionicSystemsD2.Cylinder}

We also consider the Majorana fermion of mass $m\geq 0$ on a $(1 + 1)$-dimensional flat cylinder spacetime $M = \Reals \times [ - l / 2, l / 2 )$ with spatial period $l > 0$. Specifically, in the language of \autoref{sec:ModularTheory}
we choose $\mathcal{H}_{\mathrm{r}} = \Lp2\bigl( [ - l / 2, l / 2 ), \Reals \bigr)$,  $\mathcal{L}_r = \{\psi \in \mathcal{H}_{\mathrm{r}} \mid \supp \psi \subset \mathcal{B}\}$, where again $\mathcal{B} \subset [ - l / 2, l / 2 )$ is an interval or union of intervals, and $\chi$ is the multiplication with the characteristic function of $\mathcal{B}$.

For the operator $A$, we heuristically have $A^{1/2} = (m - \partial_x) / \sqrt{-\partial_x^2 + m^2}$ as in the Minkowski case; but the differential operator now depends on a choice of boundary conditions, labelled by a parameter $\xi \in \{0, 1\}$:
periodic boundary conditions ($\xi = 0$) correspond to the Ramond sector, while antiperiodic boundary conditions ($\xi = 1$) are known as the Neveu-Schwarz (or vacuum) sector.
Specifically, see \autoref{appx:SkewSymmetricGenerator.Cylinder.Massive}, the kernel of the generator $S$ is given for $m>0$ by the modified Fourier series 
\begin{align}
\label{eq:FermionicCylinder.SkewSymmetricGenerator.Series}
    S(x, y)
  &= - \frac{2 \i}{l} \lim_{\varepsilon \to 0^+}
    \sum_{k = -\infty}^{\infty}
      \arctan\left( \frac{k + \frac{\xi}{2}}{\mu} \right)
      \e^{2 \pi \i \left( k + \frac{\xi}{2} \right) \frac{x - y}{l} - \varepsilon \abs{k}}
  \eqend{,}
\end{align}
where $\mu \coloneq \frac{m l}{2 \pi}$; note the analogy with \eqref{eq:Fermionic.SkewSymmetricGenerator.Integral}. 
In the massless limit, $m \to 0$, this formally gives
\begin{align}
\label{eq:FermionicCylinder.SkewSymmetricGenerator.MomentumSpace.Massless}
    S_0( x, y )
  &= - \frac{\pi \i}{l} \lim_{\varepsilon \to 0^+}
      \sum_{k = -\infty}^{\infty}
        \sgn\left( k + \frac{\xi}{2} \right)
        \e^{2 \pi \i \left( k + \frac{\xi}{2} \right) \frac{x - y}{l} - \varepsilon \lvert k \rvert}
  \eqend{,}
\end{align}
where the signum function takes the value 0 at 0 ($k = 0$ and $\xi = 0$); and we take this as our definition of $S$, respectively $A$, in the case $m = 0$. Note that in the case of periodic boundary conditions ($\xi = 0$), one may add a further constant to the summand $k = 0$ while retaining the skew-symmetry of $S_0$; this corresponds to a choice of ``zero mode'' and hence of different vacuum states on the massless CAR algebra, and one might even consider mixtures of these, cf.~\cite{KlichVamanWong:2017,CadamuroFroebGuillem:2024}. We will here consider only the choice \eqref{eq:FermionicCylinder.SkewSymmetricGenerator.MomentumSpace.Massless}.

The anti-locality condition \eqref{eq:OneParticleStructure.NonlocalityCondition} for the corresponding operators $A$ can be shown, for example, by a Reeh-Schlieder argument on the level of the complex Hilbert space in analogy to the Minkowski case above.

For our implementation, it is advantageous to use more explicit forms of the operator kernel of $S$,  analogous to \eqref{eq:Fermionic.SkewSymmetricGenerator.PV} in the Minkowski case, rather than the infinite Fourier series \eqref{eq:FermionicCylinder.SkewSymmetricGenerator.Series}.
For the computation of the Fourier series, one rewrites it in terms of Chebyshev polynomials and uses their properties, so that the massless skew-symmetric generator $S_0$ becomes 
\begin{align}
\label{eq:FermionicCylinder.SkewSymmetricGenerator.Series.Massless}
		S_{0}( x, y )
	&= \begin{dcases}
            \frac{ \pi}{l} \pvalue \cot\left( \pi \frac{x - y}{l} \right)
		& \text{if~} \xi = 0 \eqend{,} \\
			\frac{\pi}{l} \pvalue \csc\left( \pi \frac{x - y}{l} \right)
		& \text{if~} \xi = 1 \eqend{.}
		\end{dcases}
\end{align}
While such explicit form for $S(x,y)$ is not known for generic $m \geq 0$, we will show in~\autoref{appx:SkewSymmetricGenerator.Cylinder.Massive} that
\begin{align}
\label{eq:FermionicCylinder.SkewSymmetricGenerator.MassiveIntegral.SpecialCases}
    S(x, y)
  &= S_{0}(x, y)
    - \sgn(x - y)
    \int_{0}^{m}
      s(\tilde{m}, x - y)
    \id{\tilde{m}}
  \eqend{,}
\end{align}
with the integrand
\begin{subequations}
\eqseqlabel{eq:FermionicCylinder.SkewSymmetricGenerator.MassiveIntegral.SpecialCases.Integrands}
\begin{align}
    \xi = 0: \quad
    s(m, x - y)
  &= \frac{
      \sinh\left( m \frac{l}{2} - m \lvert x - y \rvert \right)
    }{
      \sinh\left( m \frac{l}{2} \right)
    }
  \xrightarrow[m \to 0^+]{} 1 - \frac{2 \abs{x - y}}{l}
  \eqend{,}
\nexteq
    \xi = 1: \quad
    s(m, x - y)
  &= \frac{
      \cosh\left( m \frac{l}{2} - m \lvert x - y \rvert \right)
    }{
      \cosh\left( m \frac{l}{2} \right)
    }
  \xrightarrow[m \to 0^+]{} 1
  \eqend{,}
\end{align}
\end{subequations}
for periodic and antiperiodic boundary conditions, respectively.
The formulation in terms of the smooth integrands $s(m, x - y)$ is advantageous for the numerical implementation.

\section{Numerical approach}
\label{sec:NumericalApproach}

In the following, we discuss the numerical approximation of the structures introduced in the previous section.
Essentially, we are going to choose a finite (and hence necessarily incomplete) orthonormal basis of $\mathcal{H}_{\mathrm{r}}$, see \autoref{ssec:NumericalApproach.ComplexStructure}, and approximate all operators with matrices in this basis. 
While we do not prove rigorous convergence as the number of basis elements is increased, we do offer some comments on how to compare the numerical results with the exact solution where available (\autoref{ssec:NumericalApproach.Smearing}). 
Then, we use a wedge region in $(1 + 1)$-dimensional Minkowski spacetime as a test case to show that our numerical approximations are in line with the known Bisognano--Wichmann result in that case (\autoref{ssec:NumericalApproach.RightWedge}).

\subsection{Discretisation of the complex structure}
\label{ssec:NumericalApproach.ComplexStructure}

For the $(1 + 1)$-dimensional Majorana field both on Minkowski space, $\mathcal{H}_{\mathrm{r}} = \Lp2( \Reals, \Reals )$, and on the circle, $\mathcal{H}_{\mathrm{r}} = \Lp2\bigl( [ -l/2, l/2 ), \Reals \bigr)$, we need 
to discretise the real-valued square-integrable functions over an interval $[ -b, b ] \subset \Reals$. (In the Minkowski case, a large-distance cutoff $b$ is introduced, which leads to additional numerical artefacts that are resolved for large $b$; for the cylinder, we simply have $b = l/2$.) To that end, we use box functions indexed by $i \in \{ 0, \dots, n - 1 \}$ for a given discretisation resolution $n \in \Naturals$. 
Each box function $e_i^{( n, b )}$ has support on the interval $[ a_i, b_i ]$ where $b_0 < \dots < b_{n - 1} = b$, $a_0 = -b$ and $a_i = b_{i - 1}$ for $i > 0$; in formulas,
\begin{align}
\label{eq:Discretisation.Function}
		e_i^{( n, b )}( x )
	&= n_i \upTheta( x - a_i ) \upTheta( b_i - x )
	\eqend{,}
\end{align}
where the constant $n_i \coloneq ( b_i - a_i )^{-1/2}$ makes the box functions orthonormal with respect to the $\Lp2$-inner product. 
For the implementation, we choose the same number of box functions supported in the subspace region $\mathcal{B}$ and in its complement. 
Within $\mathcal{B}$, the grid values $a_i$ are linearly spaced, while the grid spacing varies outside of $\mathcal{B}$ depending on the setting.
Specifically for the right wedge, the grid over the complement region is also linearly spaced, and for the complement of the double cone regions, the width of the box functions increases (if necessary) with increasing distance from $\mathcal{B}$.
This way of discretising the base regions is analogous to our previous work~\cite{BostelmannCadamuroMinz:2023}. 
Note that the box functions are a particularly simple choice for the discretisation, but they are not everywhere continuous. 
We also tested piecewise-linear, triangle shaped basis functions, which are continuous but complicate other parts of the approximation; they did not lead to significantly different results.
The results in the present paper make use of box functions~\eqref{eq:Discretisation.Function} only.

The matrix approximations of $A^{\pm 1/4}$, $B$ and $M_-$ are computed by means of matrix multiplication and eigenvalue decomposition from the discretisation of $\chi$ and $S$.
Since the eigenvalues of $B$ turn out to be very close to $\pm 1$, i.e., the boundary of the domain of $\artanh$, it is crucial here to use an extended floating point precision (implemented with \cite{mpmath}) with $1.5 n$ (cylinder) and $1.75 n$ decimal digits (Minkowski), which adds to the computation time required.

In the basis $\{e_i^{( n, b )}\}$, the characteristic function $\chi$ is a diagonal matrix taking the values 1 for all normalised box function within the subspace region and 0 outside.
The discretisation of the skew-symmetric generator $S$ follows from the double integrals 
\begin{subequations}
\eqseqlabel{eq:Fermionic.Majorana.ComplexStructure.Discretising}
\begin{align}
  \picklhs{
    S_{i j}^{(n, b)}
  }
  &\coloneq \innerProd{e_i^{(n, b)}}{S e_j^{(n, b)}}
\nexteq
  \skiplhs
  &= \iint_{-\infty}^{\infty}
      e_i^{(n, b)}(x) S(x,y) e_j^{(n, b)}(y)
    \id{y}
    \id{x}
\nexteq
  \skiplhs
  &= n_i n_j \int_{x = a_i}^{b_i} \int_{y = a_j}^{b_j}
      f(x - y)
	\id{y} \id{x}
  \eqend{,}
\end{align}
\end{subequations}
where $f(x - y) \coloneq S (x, y)$ is a skew-symmetric convolution kernel, hence $S_{i j}^{(n, b)} = - S_{j i}^{(n, b)}$. 
We change coordinates to the difference $t = y - x$ and sum $s = y + x$, and obtain 
\begin{align}
\label{eq:Fermionic.Majorana.ComplexStructure.Discretised}
    S_{i j}^{(n, b)}
  &= n_i n_j \Bigl( F(b_j - a_i) - F(b_j - b_i) - F(a_j - a_i) + F(a_j - b_i) \Bigr) 
\end{align}
for $i < j$ in terms of the function $F( x ) \coloneq x F_0( x ) - F_1( x )$, where
\begin{align}
\label{eq:Fermionic.Majorana.ComplexStructure.Discretised.Antiderivatives}
    F_n(x)
  &\coloneq - \int_{x}^{r} t^n f(-t) \id{t}
\end{align}
with an upper integration bound $r$, the choice of which does not change the r.h.s.\ of \eqref{eq:Fermionic.Majorana.ComplexStructure.Discretised}. 
Specifically for Minkowski spacetime and $m > 0$, we have from \eqref{eq:Fermionic.SkewSymmetricGenerator} with the choice $r \to \infty$,
\begin{align}
\label{eq:Fermionic.Majorana.ComplexStructure.Discretised.Minkowski}
    F_0(x)
  &= \upGamma(0, m x),
  &
    F_1(x)
  &= \frac{1}{m} \e^{- m x}
  \eqend{,}
\end{align}
while at $m = 0$, we take $r = 1$,
\begin{align}
\label{eq:Fermionic.Majorana.ComplexStructure.Discretised.Minkowski.Massless}
    F_{0, m = 0}(x)
  &= -\log(x),
  &
    F_{1, m = 0}(x)
  &= 1 - x
  \eqend{.}
\end{align}
For the cylinder spacetime, we treat the discretisation of the massless and massive kernels separately. 
The implementation of the massless term follows from the antiderivatives of \eqref{eq:FermionicCylinder.SkewSymmetricGenerator.Series.Massless}, choosing the upper integration bound $r = \frac{l}{2}$,   
\begin{subequations}
\eqseqlabel{eq:Fermionic.Majorana.ComplexStructure.Discretised.Cylinder.Massless}
\begin{align}
    F_{0, m = 0}(x)
  &= \begin{dcases}
      - \log\biggl\lvert \sin\left( \frac{\pi x}{l} \right) \biggr\rvert
    & \text{if~} \xi = 0
      \eqend{,}
    \\
      - \log\biggl\lvert \tan\left( \frac{\pi x}{2 l} \right) \biggr\rvert
    & \text{if~} \xi = 1
      \eqend{;}
    \end{dcases}
\nexteq
    F_{1, m = 0}(x)
  &= \begin{dcases}
      \frac{l}{\pi}
      \int_{\frac{\pi x}{l}}^{\frac{\pi}{2}}
        \vartheta \cot \vartheta
      \id{\vartheta}
    & \text{for~} \xi = 0
      \eqend{,}
    \\
      \frac{l}{\pi}
      \int_{\frac{\pi x}{l}}^{\frac{\pi}{2}}
        \vartheta \csc \vartheta
      \id{\vartheta}
    & \text{for~} \xi = 1
      \eqend{.}
    \end{dcases}
\end{align}
\end{subequations}
Analytic solutions to the integrals of $F_{1, m = 0}$ involve evaluations of the poly-logarithm function, which are numerically inefficient. 
Since the integrands are real, smooth functions without singularities in the integration domain, we use numerical integration to approximate them. 

Similarly, the functions $F_n$ for the massive correction terms in \eqref{eq:FermionicCylinder.SkewSymmetricGenerator.MassiveIntegral.SpecialCases} become double integrals over integrable functions, so that we can interchange the order of integration. 
The inner integral over $t$ has analytic solutions, thus the slow double integration reduces to a faster single integration in the numerical implementation, 
\begin{subequations}
\eqseqlabel{eq:Fermionic.Majorana.ComplexStructure.Discretised.Cylinder.Massive}
\begin{align}
    F_n(x)
  &= F_{n, m = 0}(x) + \int_{0}^{m} f_n(\tilde{m}, x) \id{\tilde{m}}
  \eqend{,}
\nexteq
    f_0(m, x)
  &= \begin{dcases}
      - \frac{
        2 \sinh^2\Bigl( \frac{m}{2} \bigl( \frac{l}{2} - x \bigr) \Bigr)
      }{
        m \sinh\Bigl( m \frac{l}{2} \Bigr)
      }
    & \text{for~} \xi = 0 \eqend{,} \\
      - \frac{
        \sinh\Bigl( m \bigl( \frac{l}{2} - x \bigr) \Bigr)
      }{
        m \cosh\Bigl( m \frac{l}{2} \Bigr)
      }
    & \text{for~} \xi = 1 \eqend{,}
    \end{dcases}
\nexteq
    f_1(m, x)
  &= \begin{dcases}
      x f_0(m, x)
      - \frac{
        \frac{1}{m} \sinh\Bigl( m \bigl( \frac{l}{2} - x \bigr) \Bigr)
        - \bigl( \frac{l}{2} - x \bigr)
      }{
        m \sinh\Bigl( m \frac{l}{2} \Bigr)
      }
    & \text{for~} \xi = 0 \eqend{,} \\
      x f_0(m, x)
      - \frac{
        2 \sinh^2\Bigl( \frac{m}{2} \bigl( \frac{l}{2} - x \bigr) \Bigr)
      }{
        m^2 \cosh\Bigl( m \frac{l}{2} \Bigr)
      }
    & \text{for~} \xi = 1 \eqend{.}
    \end{dcases}
\end{align}
\end{subequations}

\subsection{Comparing numerical results to their analytic references}
\label{ssec:NumericalApproach.Smearing}

In special cases, we want to compare our numerical results for the operators $M_-$ to known analytic expressions. Even in those, however, the kernel of $M_-$ is not a smooth function, but rather a bidistribution. We can hence only expect our numerical approximations to converge in the weak sense, i.e., for sufficiently regular test functions $h,h'$ and suitable grid points $a_i$ and $b_i$,
\begin{equation}\label{eq:weakConv}
    \sum_{k,m} \innerProd{ h}{ e_k^{(n,b)} } \left( M_-^{(n,b)} \right)_{k m} \innerProd{ e_m^{(n,b)}}{ h'} 
     \xrightarrow{n \to \infty} \iint h(x) M_-( x, y ) h'(y) \id{x} \id{y}
\end{equation}
(in the Minkowski case, $b \to \infty$ as well).
We should hence compare the left-hand side with the right-hand side of \eqref{eq:weakConv} rather than comparing the kernels pointwise. 

For the test functions $h$, $h'$, we specifically choose Gaussian-like functions with peaks at different positions $x_i$. In this way, each side of \eqref{eq:weakConv} yields a numerical matrix with an entry for each combination of peak positions $( x_i, x_j )$, which can then meaningfully be compared. 

For the case of Minkowski spacetime, we take a sequence of $\Lp2$-normalised Gaussians 
\begin{align}
\label{eq:Gaussian}
		  h_i( x )
	&= \frac{1}{\sqrt[4]{\pi \sigma^2}}
		  \exp\left(
		  - \frac{( x - x_i )^2}{2 \sigma^2}
		  \right)
\end{align}
with a standard deviation $\sigma$ that is wide enough to average over several box functions but small enough to retain details that are present in the discretised data. 

For the cylinder spacetime, we take (quasi)-periodic Gaussians, which have a simple representation for periodic ($\xi = 0$) and antiperiodic ($\xi = 1$) boundaries in terms of the third elliptic theta function~\cite[§20.2]{DLMF},
\begin{subequations}
\eqseqlabel{eq:QuasiPeriodicGaussian}
\begin{align}
	\picklhs{
		  h_i( x )
	}
	&\coloneq \frac{1}{\sqrt[4]{\pi \sigma^2}}
		\sum_{k = -\infty}^{\infty}
			(-1)^{\xi k}
			\exp\left(
				- \frac{( x - k l - x_i )^2}{2 \sigma^2}
			\right)
\nexteq
	\skiplhs
	&= \frac{1}{\sqrt[4]{\pi \sigma^2}}
		\exp\left(
			- \frac{( x - x_i )^2}{2 \sigma^2}
		\right)
		\upvartheta_3\left(
			\i \frac{( x - x_i ) l}{2 \sigma^2}
			- \frac{\pi \xi}{2}
			, \e^{-l^2/2 \sigma^2} 
		\right)
	\eqend{.}
\end{align}
\end{subequations}
Though this representation uses complex values, the test functions $h_i$ are real-valued, as required.

\subsection{Test case: the right wedge in Minkowski spacetime}
\label{ssec:NumericalApproach.RightWedge}

To demonstrate the validity of the numerical method, we consider the subspace of a right wedge in 2-dimensional Minkowski spacetime, where the analytic solution for $\Delta$ is known for any mass parameter. 
In time-0 formalism, the wedge corresponds to the region $\mathcal{B} = [ 0, \infty )$, i.e.\@, 
\begin{align}
\label{eq:RightWedge.Subspace}
		\mathcal{L}_{\mathrm{r}}
	&\coloneq \bigl\{
			f \in \mathcal{H}_{\mathrm{r}}
		\bigm\vert
			\supp f \subset [ 0, \infty )
		\bigr\}
    \subset \mathcal{H}_{\mathrm{r}} = \Lp2( \Reals, \Reals )
    \eqend{.}
\end{align}
By the theorem of Bisognano-Wichmann \cite{BisognanoWichmann:1976}, we know that the modular operator $\Delta$ is related to the representation $U$ of the Lorentz boost with the rapidity parameter $s \in \Reals$,
\begin{align}
\label{eq:RightWedge.ModularOperatorLorentzBoost}
    U(2 \pi s)
  &= \Delta^{-\i s}
  \eqend{.}
\end{align}
In our formulation, the modular Hamiltonian on the one-particle Hilbert space is therefore given by
\begin{align}
\label{eq:Fermionic.RightWedge.ModularGeneratorKernel.AtTimeZero}
    (-i_A \log \Delta)(x, y)
  &= \pi (x + y)
    \begin{pmatrix}
        0 &
        m + \partial_x \\
        - (m - \partial_x) &
        0
    \end{pmatrix}
    \updelta(x - y)
  \eqend{,}
\end{align}
as we explain in more detail in \autoref{appx:Majorana.StandardFormalism}.
The upper right block of this matrix is the kernel
\begin{align}
\label{eq:Fermionic.RightWedge.ModularGeneratorKernel.Block}
    M_-(x, y)
  &= \Bigl(
      \pi m (x + y)
      + \pi (x \partial_x - y \partial_y)
    \Bigr)
    \updelta(x - y)
  \eqend{,}
\end{align}
which uniquely splits into symmetric and skew-symmetric parts, $M_- = M_{-, {\mathrm{sym}}} + M_{-, {\mathrm{skew}}}$ with
\begin{align}
\label{eq:Fermionic.ModularGeneratorKernel.Block.Split}
    M_{-, {\mathrm{sym}}}(y, x)
  &= M_{-, {\mathrm{sym}}}(x, y)
  \eqend{,}
  &
    M_{-, {\mathrm{skew}}}(y, x)
  &= -M_{-, {\mathrm{skew}}}(x, y)
  \eqend{.}
\end{align}
The symmetric part is a mass-dependent multiplication operator, while the skew-symmetric, mass-independent term is a first-order differential operator.

\begin{figure}
    \centering
    \begin{subfigure}{0.95\textwidth}
        \centering
        \includegraphics{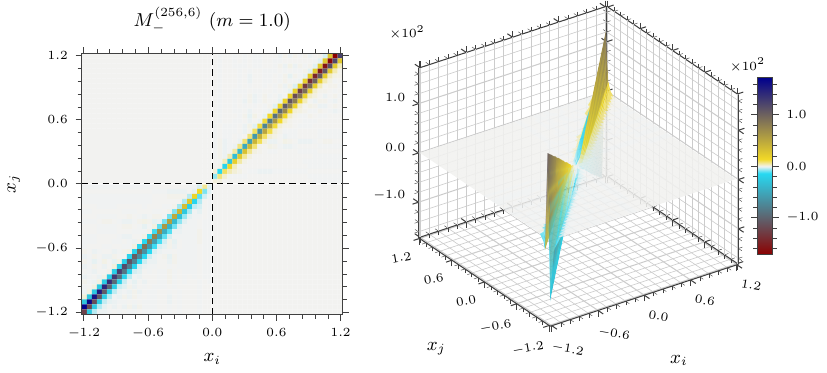}
        \caption{\label{fig:RightWedge.m1.0} Numeric result for the block $M_-$ (mass $m = 1.0$). Dashed lines indicate the subspace boundary.}
    \end{subfigure}
    \\[1em]
    \begin{subfigure}{0.95\textwidth}
        \centering
        \includegraphics{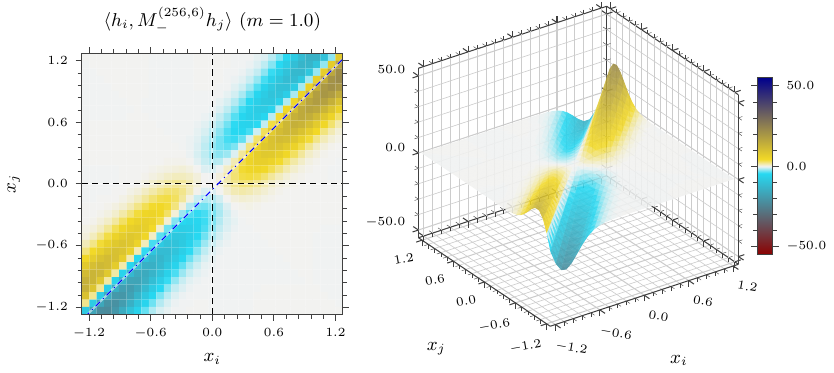}
        \caption{\label{fig:RightWedge.m1.0.Smeared} The result~\autoref{fig:RightWedge.m1.0} integrated against Gaussian test functions at various positions $x_i$, $x_j$.}
    \end{subfigure}
    \\[1em]
    \begin{subfigure}{0.95\textwidth}
        \centering
        \includegraphics{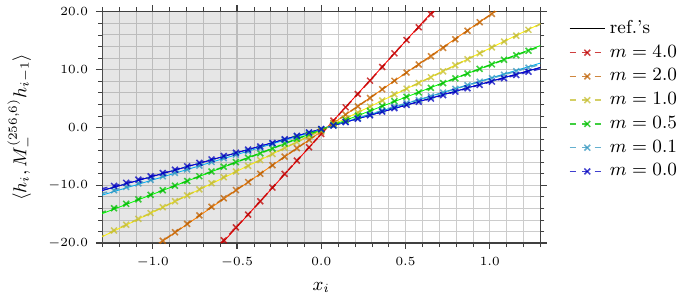}
        \caption{\label{fig:RightWedge.Diagonal} Comparison of the numeric results (dashed lines, cross marks) against the analytic references (solid lines) parallel to the diagonal for different mass parameters.}
    \end{subfigure}
    \caption{\label{fig:RightWedge} Processing steps for a comparison of our numeric method against analytic solutions (from top down): First, we compute $M_-$ in a fixed discretisation; second, we integrate it against Gaussian test functions; third, we repeat and collect results along a diagonal (blue dash-dot line) for different masses.}
\end{figure}

For the numerical approximation $M_-^{( n, b )}$ computed as described in \autoref{ssec:NumericalApproach.ComplexStructure}, we choose a cutoff $b = 6$ and a resolution of $n = 256$ box functions. 
We obtain the numeric result shown in~\autoref{fig:RightWedge.m1.0} for mass $m = 1$. 
In agreement with the analytic expression~\eqref{eq:Fermionic.RightWedge.ModularGeneratorKernel.Block}, the numeric result is concentrated along the diagonal $x = y$, and the main and secondary diagonals are opposite in sign, numerically approximating the differential operator. 
For a more detailed comparison, we integrate against test functions.  

As test functions $h_i$, we take the Gaussians \eqref{eq:Gaussian} with $\sigma = 0.163$, which is  about 3.5 times wider than the box function width $2 b / n$ in the discretisation, and the peaks of the Gaussians are centered at equally spaced positions $x_i$. 
Integrating the operator kernel against these test functions, we get the symmetric and skew-symmetric matrix elements 
\begin{subequations}
\eqseqlabel{eq:Fermionic.RightWedge.ModularGeneratorKernel.Block.Smeared.Parts}
\begin{align}
    \innerProd{h_i}{M_{-, {\mathrm{sym}}} h_j}
  &= \pi m (x_i + x_j)
    \exp\left( - \frac{(x_i - x_j)^2}{4 \sigma^2} \right)
  \eqend{,}
\nexteq
    \innerProd{h_i}{M_{-, \mathrm{skew}} h_j}
  &= - \pi \frac{x_i^2 - x_j^2}{2 \sigma^2}
    \exp\left( - \frac{(x_i - x_j)^2}{4 \sigma^2} \right)
  \eqend{.}
\end{align}
\end{subequations}
Summing over the numeric data with the discretised versions of the test functions
\begin{align}
\label{eq:TestFunction.Discretized}
    h_i^{(n, b)}(x)
  &= \sum_{k = 0}^{n - 1} \innerProd{e_k^{(n, b)}}{h_i} e_k^{(n, b)}(x)
\end{align}
yields the corresponding numeric data shown in~\autoref{fig:RightWedge.m1.0.Smeared} (where we use the compact notation $\langle h_i, M_-^{(n, b)} h_j \rangle$ to denote the matrix elements computed with \eqref{eq:TestFunction.Discretized}). 
For comparison, we plot the analytic results parallel to the matrix diagonal (dash-dot line in~\autoref{fig:RightWedge.m1.0.Smeared}) at $i = j + 1$ with solid lines and overlay the corresponding diagonal of the numeric data for different mass parameter $m$ in~\autoref{fig:RightWedge.Diagonal}.
This shows good agreement between the numerical results and analytic predictions.

\section{Numerical results}
\label{sec:NumericalResults}

We now proceed to our numerical results for the Majorana field in cases where an analytic expression for the modular generator is \emph{not} known: Specifically, we consider one and two double cones as a subregion of the cylinder in the case of generic mass, $m \geq 0$. Only for the case $m=0$, explicit expressions for $\log\Delta$ have been computed~\cite{KlichVamanWong:2017}; translated into our framework (see \autoref{appx:Cylinder.MasslessModularHamiltonian}), these results can serve as a point of comparison for the numerical data.

In \autoref{ssec:CylinderResults.DoubleCone}, we first present results for an individual double cone on a cylinder with antiperiodic boundary conditions.
Then, in \autoref{ssec:Results.DoubleCone.BoundaryConditions}, we discuss the influence of the choice of boundary conditions, as well as links to the situation on Minkowski spacetime and the effect of the large-distance cutoff there.
In \autoref{ssec:CylinderResults.DoubleCone2}, we consider a subregion of two double cones within the cylinder.

\subsection{Results: Mass-dependence of the modular generator for a double cone}
\label{ssec:CylinderResults.DoubleCone}

We first consider the modular Hamiltonian of a double cone on a cylinder spacetime; that is, the region considered is an interval $\mathcal{B} = [-w/2, w/2]$ on the circle with period $l$ with $w < l$, for example $w = l/2$.

As a reference, in the case $m = 0$, the modular Hamiltonian for periodic ($\xi = 0$) or antiperiodic ($\xi = 1$) boundary conditions has the following form (see \autoref{appx:Cylinder.MasslessModularHamiltonian})
\begin{align}
\label{eq:FermionicCylinder.OneDoubleCone.ModularHamiltonianBlock}
    M_{-}(x, y)
  &\pickindent{=}
    \pi \left( z'(x)^{-1} + z'(y)^{-1} \right)
    \updelta'(x - y)
    + \updelta_{\xi0} \frac{\pi^2}{l}
    \left( z'(x)^{-1} + z'(y)^{-1} \right)
    \updelta(x + y)
  \eqend{}
\end{align}
with the profile function following from \cite[Eq.~13]{KlichVamanWong:2017},
\begin{align}
\label{eq:FermionicCylinder.OneDoubleCone.MulticomponentFactor}
    z'(x)^{-1}
  &= \frac{l}{2 \pi}
    \csc\left( \frac{\pi w}{l} \right)
    \left(
      \cos\left( \frac{2 \pi x}{l} \right)
      - \cos\left( \frac{\pi w}{l} \right)
    \right)
  \eqend{.}
\end{align}
The massless kernel \eqref{eq:FermionicCylinder.OneDoubleCone.ModularHamiltonianBlock} follows as a special case of the modular operator restricted to a multi-interval subspace \cite{KlichVamanWong:2017}.
Note that \cite{KlichVamanWong:2017} consider only the modular Hamiltonian restricted to the subspace region.
Since the complement of an interval on the circle is also an interval, the restricted expressions actually extends to the full Hilbert space $\mathcal{H}_{\mathrm{r}} = \Lp2\bigl( [ -l/2, l/2 ), \Reals \bigr)$ (and similar for any disjoint union of intervals).
Thus we can compare the massless solution as a reference to the numeric data, which is approximating the modular general on the entire space.
The first term of $M_-$ is skew-symmetric in $x$ and $y$, and the second term (the massless ``zero mode'') is only present in case of periodicity $\xi = 0$.
For antiperiodic boundaries ($\xi = 1$), the symmetric part $M_{-, \mathrm{sym}}(x, y)$ vanishes and symmetric contributions may only appear for massive cases $m > 0$.
We will therefore consider the case $\xi = 1$ for the remainder of the subsection.

\begin{figure}
  \centering
  \includegraphics{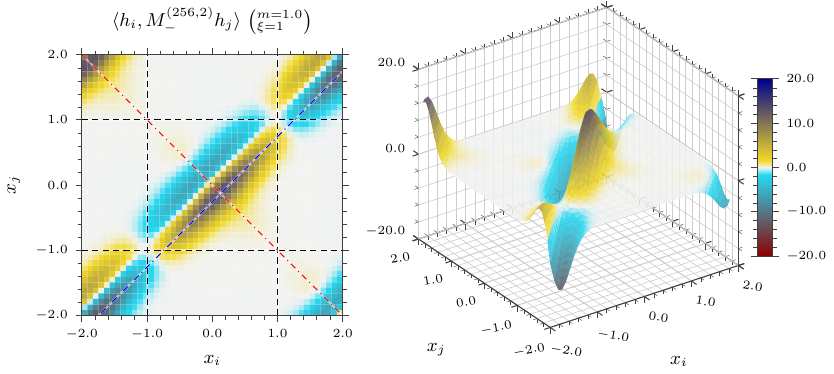}
  \caption{\label{fig:CylinderDoubleCone.xi1.m1.0} Numerical result for the block $M_-$ of the modular generator for mass $m = 1.0$ and antiperiodic boundary conditions ($\xi = 1$). }
\end{figure}

In~\autoref{fig:CylinderDoubleCone.xi1.m1.0}, we show the numeric results for a massive field ($m = 1$) and the double cone over an interval half the width of an antiperiodic circle ($w = 2$, $l = 4$, $\xi = 1$) so that the discretising box functions all have the same width inside and outside the subspace region. 
The results are shown after integrating against the quasiperiodic Gaussian test functions \eqref{eq:QuasiPeriodicGaussian}.

\begin{figure}
  \centering
  \includegraphics{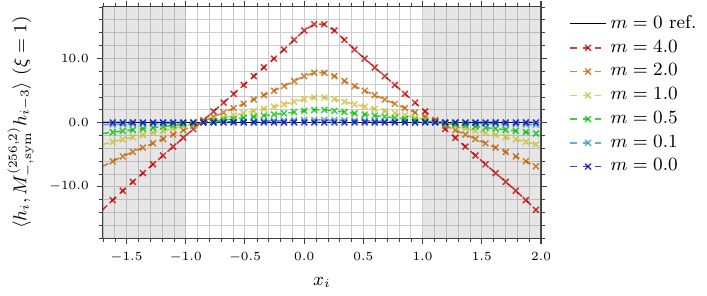}
  \caption{\label{fig:CylinderDoubleCone.xi1.SymmetricDiagonal} The symmetric part of the modular generator for a double cone on the antiperiodic cylinder ($\xi = 1$) for different masses. Results parallel to the diagonal (blue dash-dot line in \autoref{fig:CylinderDoubleCone.xi1.m1.0}). of the symmetric component for different masses.}
\end{figure}

For a more in-depth analysis of the mass dependence, we split the numeric results into symmetric and skew-symmetric parts and first consider the symmetric part of the kernel along a parallel to the diagonal (blue dash-dot line in~\autoref{fig:CylinderDoubleCone.xi1.m1.0}).
The results are shown in \autoref{fig:CylinderDoubleCone.xi1.SymmetricDiagonal};
the curves show nearly linear slopes (from the edges of the region towards the centre) approximately proportional to the mass parameter.
In this respect, the behaviour is similar to the symmetric part $M_{-, \mathrm{sym}}$ in the case of the wedge subspace in Minkowski spacetime \eqref{eq:Fermionic.RightWedge.ModularGeneratorKernel.Block.Smeared.Parts}. 

\begin{figure}[t]
  \centering
  \begin{subfigure}{0.95\textwidth}
      \centering
      \includegraphics{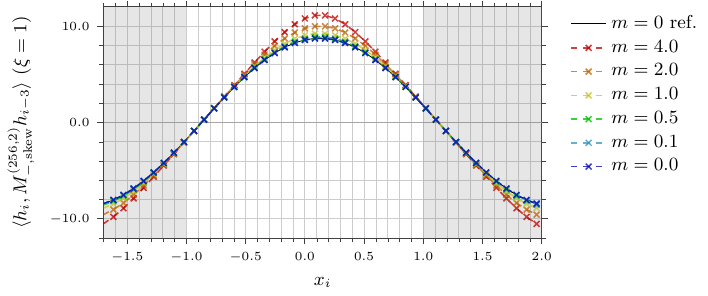}
      \caption{\label{fig:CylinderDoubleCone.xi1.SkewSymmetricDiagonal} Parallel to the diagonal (blue dash-dot line in \autoref{fig:CylinderDoubleCone.xi1.m1.0}).}
  \end{subfigure}
  \\[1em]
  \begin{subfigure}{0.95\textwidth}
      \centering
      \includegraphics{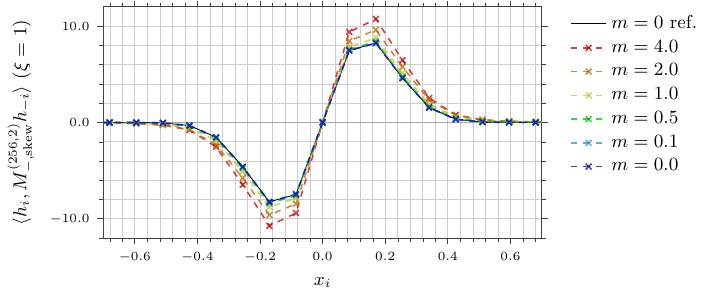}
      \caption{\label{fig:CylinderDoubleCone.xi1.SkewSymmetricAntidiagonal} Along the antidiagonal (red dash-dot line in \autoref{fig:CylinderDoubleCone.xi1.m1.0}).}
  \end{subfigure}
  \caption{\label{fig:CylinderDoubleCone.xi1.Skewsymmetric} The skew-symmetric part of the modular generator of a double cone on the anti-periodic cylinder ($\xi = 1$) for different masses.}
\end{figure}

On the other hand, contrary to the right wedge case, the skew-symmetric part $M_{-,\mathrm{skew}}$ also depends on the mass, though the dependence is much less significant.
This is shown in \autoref{fig:CylinderDoubleCone.xi1.Skewsymmetric} where the kernel values are plotted on the parallel to the diagonal as well as on the antidiagonal.
The behaviour along the antidiagonal suggests that the skew-symmetric part is approximately a first order differential operator like in the massless case, see \eqref{eq:FermionicCylinder.OneDoubleCone.ModularHamiltonianBlock}.

In all cases considered, we find good agreement between the exactly known massless case \eqref{eq:FermionicCylinder.OneDoubleCone.ModularHamiltonianBlock} (black solid lines) and our numerical results at low $m$ (blue dashed lines with cross marks).

\subsection{Results: Influence of boundary conditions and discretisation cutoffs}
\label{ssec:Results.DoubleCone.BoundaryConditions}

\begin{figure}
    \centering
    \begin{subfigure}{0.95\textwidth}
        \centering
        \includegraphics{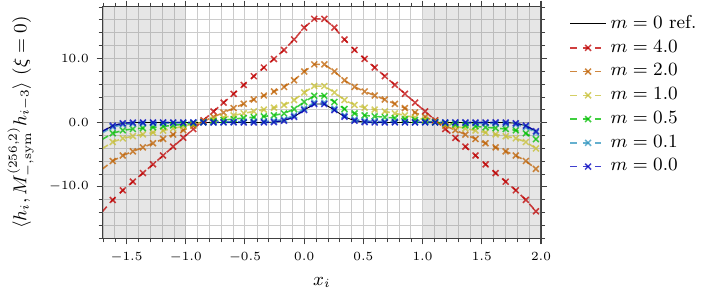}
        \caption{\label{fig:CylinderDoubleCone.xi0.SymmetricDiagonal} Parallel to the diagonal (analogous to the blue dash-dot line in \autoref{fig:CylinderDoubleCone.xi1.m1.0}) of the symmetric component.}
    \end{subfigure}
    \\[1em]
    \begin{subfigure}{0.95\textwidth}
        \centering
        \includegraphics{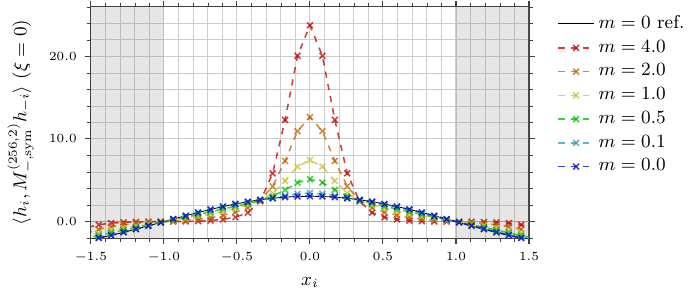}
        \caption{\label{fig:CylinderDoubleCone.xi0.SymmetricAntidiagonal} Along the antidiagonal (red dash-dot line in \autoref{fig:CylinderDoubleCone.xi1.m1.0}) of the symmetric component.}
    \end{subfigure}
    \caption{\label{fig:CylinderDoubleCone.xi0} Comparison of the modular generator on the periodic cylinder ($\xi = 0$) for different masses (and the massless reference) parallel to the diagonal and along the antidiagonal, similar to the setup shown with the blue and red dash-dot lines in \autoref{fig:CylinderDoubleCone.xi1.m1.0}, respectively. Both plots show the operator part that is symmetric in $x_i$ and $x_j$.}
\end{figure}

In the case of a double cone, we now proceed to periodic ($\xi = 0$) boundary conditions.
Compared with the antiperiodic case ($\xi = 1$), already the massless case \eqref{eq:FermionicCylinder.OneDoubleCone.ModularHamiltonianBlock} contains an extra ``zero mode'' contribution.
This is reflected in our numerical results in \autoref{fig:CylinderDoubleCone.xi0}, where the symmetric part of the kernel is shown for $\xi = 0$ and various masses, both on the parallel to the diagonal and on the antidiagonal.
The extra contribution appears to influence mostly the values near the center of the interval, and  contributes less with increasing mass -- compare \autoref{fig:CylinderDoubleCone.xi0.SymmetricDiagonal} with \autoref{fig:CylinderDoubleCone.xi1.SymmetricDiagonal}.
This may be expected since the ``zero mode'' ambiguity in the vacuum state should be an effect of the low-mass limit.

\begin{figure}
    \centering
    \includegraphics{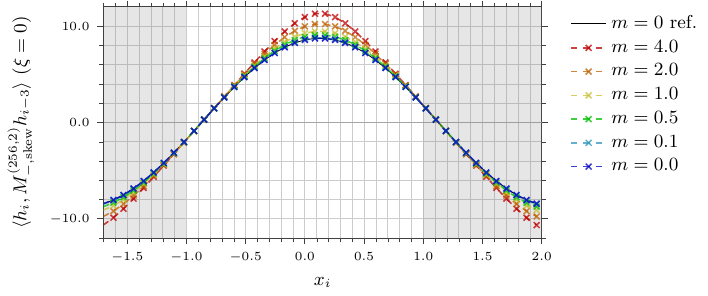}
    \caption{\label{fig:CylinderDoubleCone.xi0.SkewSymmetricDiagonal} The skew-symmetric component of $M_-$ for a double cone on the periodic cylinder ($\xi=0$) parallel to the diagonal (blue dash-dot line in \autoref{fig:CylinderDoubleCone.xi1.m1.0}) }
\end{figure}

\begin{figure}
    \centering
    \begin{subfigure}{0.95\textwidth}
        \centering
        \includegraphics{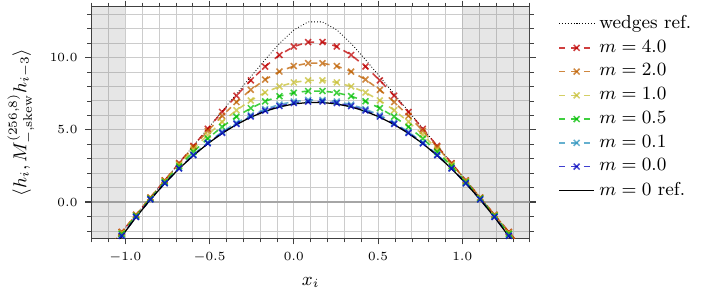}
        \caption{\label{fig:DoubleCone.b8.SkewSymmetricDiagonal} Cutoff boundary at $b = 8$.}
    \end{subfigure}
    \\[1em]
    \begin{subfigure}{0.95\textwidth}
        \centering
        \includegraphics{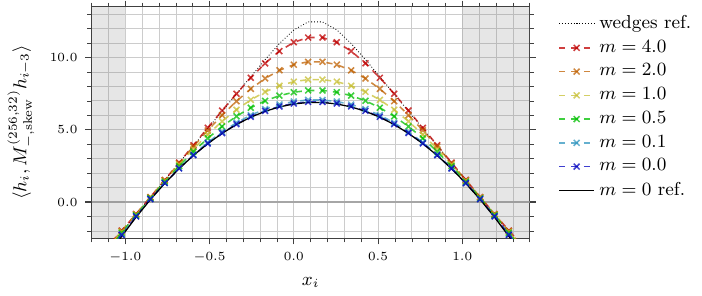}
        \caption{\label{fig:DoubleCone.b32.SkewSymmetricDiagonal} Cutoff boundary at $b = 32$.}
    \end{subfigure}
    \caption{\label{fig:DoubleCone.BoundaryIndependence} Skew-symmetric component of the modular generator for a double cone on Minkowski space, parallel to the diagonal (blue dash-dot line in \autoref{fig:CylinderDoubleCone.xi1.m1.0}), with different cutoff boundaries.}
\end{figure}

On the other hand, the skew-symmetric part of the kernel shows little dependence on the choice of boundary conditions, see \autoref{fig:CylinderDoubleCone.xi0.SkewSymmetricDiagonal} for $\xi=0$ and \autoref{fig:CylinderDoubleCone.xi1.SkewSymmetricDiagonal} for $\xi=1$.
The fact that the skew-symmetric part is almost independent of the choice of boundary conditions on the cylinder spacetime suggests that the same may hold true for the cutoff boundaries that are introduced in the numerical scheme when approximating the modular operator for an interval in Minkowski spacetime.
Indeed, when repeating the numerical computation for this case, the cutoff boundaries appear to influence primarily the symmetric part of the operator (not shown here) and leave the skew-symmetric operator part almost unaffected by the position of the cutoff $b$ (see \autoref{fig:DoubleCone.BoundaryIndependence}, where we choose two different cutoffs, $b=8$ and $b=32$). 
This also agrees well with the known exact results in the case $m = 0$ on Minkowski spacetime without boundary cutoff \cite{HislopLongo:1982,CasiniHuerta:2009}: in our notation, and for a double cone over the interval $\mathcal{B} = [ -w/2, w/2 ]$ this is the same as the formal limit $l \to \infty$ of \eqref{eq:FermionicCylinder.OneDoubleCone.ModularHamiltonianBlock} with a massless profile function
\begin{align}
\label{eq:Fermionic.DoubleCone.MasslessProfile}
    z'(x)^{-1}
  &= \frac{\left( \frac{w}{2} \right)^2 - x^2}{w}
  \eqend{.}
\end{align}

The double cone region in Minkowski spacetime is the intersection of a right wedge at $-w/2$ and a left wedge at $w/2$, so that the modular generator is bounded by these operators, $-\log \Delta \leq -\log \Delta_{\mathrm{L,R}}$, which we include in \autoref{fig:DoubleCone.BoundaryIndependence} as a reference (black dotted lines) with the combined profile function
\begin{align}
\label{eq:Fermionic.DoubleCone.UpperBoundProfile}
    z'(x)^{-1}
  &= \min\left( x + \frac{w}{2}, -x + \frac{w}{2} \right)
  \eqend{.}
\end{align}
Since in the wedge case, the skew-symmetric part of $M_-$ as in \eqref{eq:Fermionic.RightWedge.ModularGeneratorKernel.Block} is independent of the mass, the profile function \eqref{eq:Fermionic.DoubleCone.UpperBoundProfile} is an upper bound for the skew-symmetric part of the modular generator at any mass, which agrees with the numeric results.

\subsection{Results: Mass-dependence of the modular generator for two double cones}
\label{ssec:CylinderResults.DoubleCone2}

Finally, we want to show results of the modular operator for two double cones over the subspace region $\mathcal{B} = \left[ a_1, b_1 \right] \cup \left[ a_2, b_2 \right]$ on a cylinder spacetime with antiperiodic boundary conditions. 
In the massless case, the $M_-$ block within the subspace region has the integral kernel (see \autoref{appx:Cylinder.MasslessModularHamiltonian}) %
\begin{align}
\label{eq:FermionicCylinder.TwoDoubleCones.ModularHamiltonianBlock}
    M_{-}( x, y )
  &\pickindent{=}
    \pi \left( z'(x)^{-1} + z'(y)^{-1} \right) \updelta'(x - y)
  \eqbreakr
    - \frac{2 \pi^2}{l}
    \csc\left( \pi \frac{x - y}{l} \right)
    z'(y)^{-1} \updelta\bigl( v(x) - y \bigr)
\end{align}
where $v(x)$ is the non-trivial solution of $z(x) = z(y)$. 
In particular, if the two intervals have the same width and separation, i.e., $\mathcal{B} = \left[ -3l/8, -l/8 \right] \cup \left[ l/8, 3l/8 \right]$, we have
\begin{align}
\label{eq:FermionicCylinder.TwoDoubleCones.SecondSolution.SymmetricCase}
    v(x)
  &= x + \frac{l}{2}
\end{align}
modulo a multiple of $l$ due to periodicity, and the profile function takes the form 
\begin{align}
\label{eq:FermionicCylinder.TwoDoubleCones.MulticomponentFactor}
    z'(x)^{-1}
  &= -\frac{l}{4 \pi} \cos\left( \frac{4 \pi x}{l} \right)
  \eqend{.}
\end{align}
Note that for values $x \in \mathcal{B}$, $z'(x)^{-1}$ is positive, and $z'(x)^{-1} = z'\bigl( v(x) \bigr)^{-1}$.
For all $x \in [- l/2, l/2] \setminus \mathcal{B}$, the profile~\eqref{eq:FermionicCylinder.TwoDoubleCones.MulticomponentFactor} is negative.
The second term in \eqref{eq:FermionicCylinder.TwoDoubleCones.ModularHamiltonianBlock} correlates field values at $x$ with field values at $v( x )$, referred to as cross-correlation in the following. 

\begin{figure}
    \centering
    \includegraphics{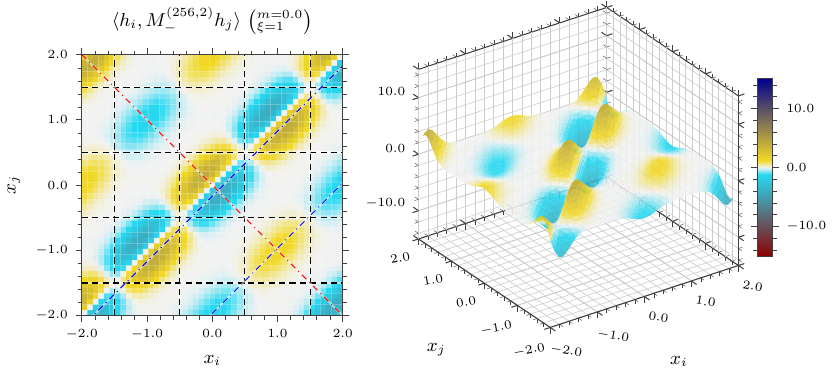}
    \caption{\label{fig:CylinderDoubleCone2.xi1.m1.0} 
    Modular generator $M_-$ for two double cones in a cylinder spacetime $(\xi = 1, m = 1)$.
    }
\end{figure}

\begin{figure}
    \centering
    \begin{subfigure}{0.95\textwidth}
        \centering
        \includegraphics{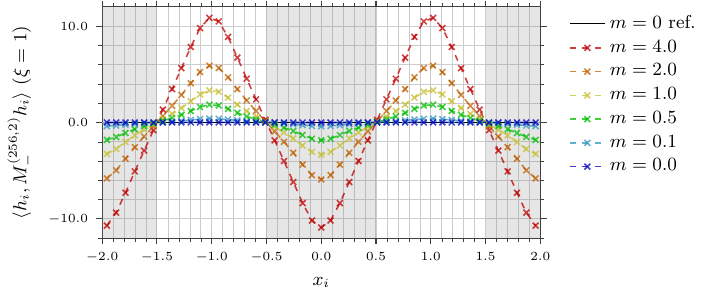}
        \caption{\label{fig:CylinderDoubleCone2.xi1.Diagonal} Parallel to the diagonal (long blue dash-dot line).}
    \end{subfigure}
    \\[1em]
    \begin{subfigure}{0.95\textwidth}
        \centering
        \includegraphics{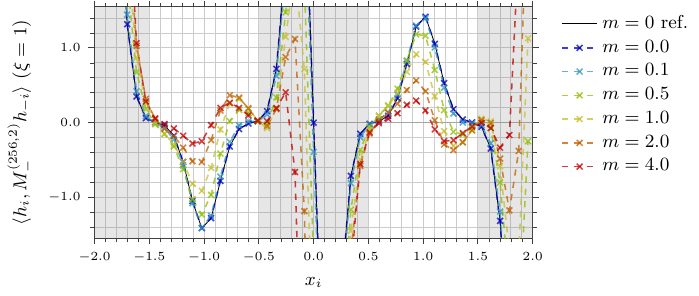}
        \caption{\label{fig:CylinderDoubleCone2.xi1.Antidiagonal} Along the antidiagonal (red dash-dot line).}
    \end{subfigure}
    \\[1em]
    \begin{subfigure}{0.95\textwidth}
        \centering
        \includegraphics{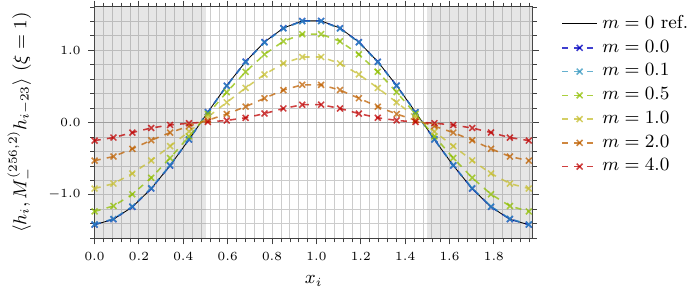}
        \caption{\label{fig:CylinderDoubleCone2.xi1.CrossCorrelationDiagonal} Along a diagonal through one of the cross-correlation regions (short blue dash-dot line).}
    \end{subfigure}
    \caption{\label{fig:CylinderDoubleCone2.CrossCorrelation} 
    Modular generator for two intervals in a cylinder spacetime for different masses, along various lines indicated in \autoref{fig:CylinderDoubleCone2.xi1.m1.0}.}
\end{figure}

The numerical results at mass $m = 1.0$ are shown in~\autoref{fig:CylinderDoubleCone2.xi1.m1.0}, where the cross-correlation terms (for $x_i \in [-3/2, -1/2]$, $x_j \in [1/2, 3/2]$, and vice versa) are clearly visible.
Note that we did not split the kernel into symmetric and skew-symmetric contributions for the present case.
A comparison for different masses is shown in \autoref{fig:CylinderDoubleCone2.CrossCorrelation}, displaying the kernel along certain lines in the square $[-2,2]^2$ as indicated by the dash-dot lines in~\autoref{fig:CylinderDoubleCone2.xi1.m1.0}.

Near to the main diagonal (\autoref{fig:CylinderDoubleCone2.xi1.Diagonal}, along the long blue dash-dot line in \autoref{fig:CylinderDoubleCone2.xi1.m1.0}), we see an increase in magnitude with the mass parameter, similar to the case of an individual double cone.
The profile of the curves for each double cone is somewhat similar to the case of an individual double cone.

Plotting the data along the antidiagonal (red dash-dot line) yields the mass-dependence of the cross-correlation term, see~\autoref{fig:CylinderDoubleCone2.xi1.Antidiagonal}. 
The profile of the curve changes from being symmetric with respect to the centre of each subspace interval in the massless case, becoming more and more skew-symmetric and decreasing in magnitude when approaching very large masses.
The values along a diagonal close to $y = v( x )$ (short blue dash-dot line in \autoref{fig:CylinderDoubleCone2.xi1.m1.0}), shown in~\autoref{fig:CylinderDoubleCone2.xi1.CrossCorrelationDiagonal}, give another indication that the  cross-correlation terms are decreasing at large masses. 
A heuristic explanation for this behaviour is that the more massive the fermions become, the shorter their correlation range is, making the two double cone regions more and more ``independent''.

\section{Conclusions}
\label{sec:Conclusion}

In this paper, we have computed a numerical approximation to the modular Hamiltonian for one and two double cones in the case of the free massive Majorana field in two spacetime dimensions.
Specifically, we considered a cylinder spacetime to avoid boundary effects from discretization cutoffs that we observed in Minkowski space. 
For the massless cylinder spacetime, we considered antiperiodic and periodic boundary conditions. 
The modular data on the level of the one-particle structure is discretised with a suitable basis of functions in position space reducing the problem to linear algebra, so that components of the modular operator like $M_-$, see~\eqref{eq:Fermionic.ModularOperator.Blocks}, are approximated by matrices. 

Our results indicate that in the case of a double cone on the antiperiodic cylinder, the symmetric part of the modular generator ($M_-$) has a local contribution (i.e., it is concentrated along the diagonal) that is mass-dependent.
This term is only present for massive particles and its magnitude is approximately proportional to the mass. 
The skew-symmetric part of the kernel of $M_-$ appears to arise from a first order derivative (like in the massless case) and is only mildly mass-dependent.

Perturbation theory in the analogous Minkowski space setup indicates that the modular Hamiltonian may also have a non-local contribution at first order in the mass $m$ \cite{CadamuroFroebMinz:2025}.
Our numerical accuracy is currently not high enough to confirm the existence of this small effect.

In the case of the \emph{periodic} cylinder, the symmetric part of $M_-$ is nonvanishing already in the massless case (``zero mode'' contribution), and our numerical results show a mass-dependence; at large masses, the periodic and antiperiodic cases differ only slightly. For the skew-symmetric part of $M_-$, the periodicity condition has little influence across the entire range of masses. Overall, we find a near-independence of $M_-$ on the boundary conditions at large masses: the influence of the ``zero mode'' is restricted to the low-mass regime. 

Similarly, for the skew-symmetric part of $M_-$ in the case of a double cone subregion of Minkowski spacetime, we find that the cutoff boundaries introduced as part of the numerical algorithm have negligible influence, as long as they are large enough.
These results agree with similar numerical studies on lattices \cite{EislerPeschel:2009,EislerTonniPeschel:2019,EislerTonniPeschel:2022}.

For two double cones within the cylinder, we find that the modular generator does not only have local components, i.e., concentrated along the diagonal, but additional ``bilocal'' terms supported in the cross-correlation regions show up in the numerical results; this is expected from the exact result in the  massless case. These bilocal terms, however, decrease in magnitude as the mass is increased. In a heuristic interpretation, heavier fermions have a shorter correlation length, so that the far-range correlation showing up in the cross-correlation regions is weaker. 

While we have verified the quality of our approximations by comparing with the known analytic result for the massless cases, and also with the wedge case in Minkowski spacetime, we do not have any rigorous error estimates on our numerics. 
Also, we did not present any proof of convergence of the finite-dimensional modular data to those in the continuum. 
This requires further work.

We restricted our attention to the $(1 + 1)$-dimensional Majorana field, which we believe yields insight into the typical structure of mass-dependence of the modular generator. However, the framework we described in \autoref{sec:ModularTheory} allows for other situations as well. For example, for the Dirac field in physical spacetime dimensions, the initial spinor data splits into two spinor parts (particles and antiparticles), and each part is described by the real Hilbert space $\mathcal{H}_{\mathrm{r}} \coloneq \Lp2( \Reals^3, \Reals ) \otimes \Reals^2$. The configuration space kernel of the operator $A^{\pm 1/2}$ in this case is, when written in terms of the Pauli matrix basis $( \one, \sigma^1, \sigma^2, \sigma^3 )$,
\begin{align}
\label{eq:Fermionic4D.ComplexStructureKernels.MomentumSpace}
    A^{\pm \frac{1}{2}}(\mathvec{x}, \mathvec{y})
  &= \frac{1}{2 \pi}
    \lim_{\varepsilon \to 0^+}\int_{-\infty}^{\infty}
      \frac{m \one \mp \i \sigma^i p_i}{\sqrt{\mathvec{p}^2 + m^2}}
      \,\e^{\i \mathvec{p} \cdot (\mathvec{x} - \mathvec{y}) - \varepsilon \abs{\mathvec{p}}}
      \id[3]{p}
  \eqend{.}
\end{align}
However, working with the discretization of these objects as functions of 3 variables would require vastly more computational power than our $(1 + 1)$-dimensional example; if the region in question has additional symmetries, such as a double cone, then a result may be obtainable (see the bosonic analogue in \cite[Sec.~6]{BostelmannCadamuroMinz:2023}), but a result for more general regions would require a substantial improvement in the numerical methods.

Likewise, the same methods developed in this paper should apply to other spacetime regions in lower dimensions, to other wave operators (e.g., linear fields coupled to an external potential), and by purification methods also to non-pure quasifree states of ferminonic systems, such as thermal states of QFT or those in linear quantum fields on a gravitational background, the complexity of the numerical evaluation again being a potential limitation.

\section*{Acknowledgements}

D.C.\ and C.M.\ are supported by the Deutsche Forschungsgemeinschaft (DFG) within the Emmy Noether grant CA1850/1-1.
C.M.~thanks Markus Fr\"ob for discussions on the computation of the analytic references and the ``zero modes'' on the periodic cylinder spacetime.

\appendix

\section{Derivation of the complex structure from the Majorana Lagrangian}
\label{appx:Majorana.StandardFormalism}

In this appendix, we show how our expressions for the modular generator and the complex structure relate to the more standard formalism in terms of fields, see \cite{FreedmanVanProeyen:2012}, in particular Sec.~I.3 there for Majorana fields in two dimensions.
The Lagrangian density for free Majorana fermion spinors $\psi$ (with Grassmann-valued components $\psi_a$, $a \in \{ 1, 2 \}$ and the Dirac adjoint $\bar{\psi} = \psi^\transpose \gamma^{0}$) in two-dimensional Minkowski spacetime (with signature $+-$) is 
\begin{align}
\label{eq:Fermionic.Majorana.Lagragian}
		  \mathscr{L}
	&= \frac{\i}{4} \left(
		    \bar{\psi} \gamma^{a} ( \nabla_{a} \psi )
		  - ( \nabla_{a} \bar{\psi} ) \gamma^{a} \psi 
		  \right)
		- \frac{1}{2} m \bar{\psi} \psi
	\eqend{,}
\end{align}
with the gamma matrices (which can be written in terms of the Pauli matrices $\sigma^i$)
\begin{subequations}
\eqseqlabel{eq:Fermionic.Majorana.GammaMatrices}
\begin{align}
    \gamma^{0}
  = \begin{pmatrix}
      0 & -\i \\
      \i & 0
    \end{pmatrix}
  = \sigma^2
  \eqend{,}
  \qquad\qquad
    \gamma^{1}
  = \begin{pmatrix}
      0 & \i \\
      \i & 0
    \end{pmatrix}
  = \i \sigma^1
  \eqend{,}
\nexteq
    \gamma^{0} \gamma^{0}
  = \one
  \eqend{,}
  \qquad
    \gamma^{0} \gamma^{1}
  = - \gamma^{1} \gamma^{0}
  = \begin{pmatrix}
      1 &   0 \\
      0 & - 1
    \end{pmatrix}
  = \sigma^3
  \eqend{,}
  \qquad
    \gamma^{1} \gamma^{1}
  &= -\one
  \eqend{.}
\end{align}
\end{subequations}
In order to obtain the block antidiagonal form of the complex structure, we transform to the spinor components $\psi_\pm \coloneq ( \psi_1 \mp \psi_2 ) / \sqrt{2}$ that fulfill the field equations 
\begin{align}
\label{eq:Fermionic.Majorana.EOM}
		  \partial_t \psi_{\pm}
	&= ( - \partial_x \mp m ) \psi_{\mp}
	\eqend{,}
\end{align}
and equal-time canonical anti-commutation relations 
\begin{align}
\label{eq:Fermionic.Majorana.Anticommutator}
		  \acomm{\psi_{\pm}( t, x )}{\psi_{\pm}( t, x' )}
	&= \updelta( x - x' ) \one
	\eqend{,}
&
		  \acomm{\psi_{\pm}( t, x )}{\psi_{\mp}( t, x' )}
	&= 0
	\eqend{.}
\end{align}
On the mass shell ($\omega = \sqrt{p^2 + m^2}$), the canonically quantized fields read 
\begin{subequations}
\eqseqlabel{eq:Fermionic.Majorana.SpinorComponentsQuantized}
\begin{align}
		  \psi_{\pm}( t, x )
	&= \frac{1}{2 \pi}
		  \int \frac{1}{\sqrt{2}} \left(
		    \rho_{\pm}( p )
		    c( p ) \e^{-\i \omega t + \i p x}
		  + \rho_{\mp}( p )
		    c^\dagger( p ) \e^{\i \omega t - \i p x}
		  \right) \id{p}
	\eqend{,}
\nexteq
		  \rho_{\pm}( p )
	&\coloneq \sqrt{\frac{\omega + p}{2 \omega}}
		\mp \i \sqrt{\frac{\omega - p}{2 \omega}}
	\eqend{,}
\end{align}
\end{subequations}
given in terms of the annihilation and creation operators $c( p )$ and $c^\dagger( p )$ that fulfill the anti-commutation relations 
\begin{align}
\label{eq:Fermionic.Majorana.CreationAnnihilationAnticommutator}
		  \acomm{c(p)}{c(p')}
	&= 0
	\eqend{,}
&
		  \acomm{c(p)}{c^\dagger(p')}
	&= 2 \pi \updelta(p - p') \one
	\eqend{.}
\end{align}

The one-particle Hilbert space contains all real-valued functions that describe the two components of initial data, $f = f_+ \oplus f_- \in \mathcal{H}_{\mathrm{r}} \oplus \mathcal{H}_{\mathrm{r}} = \mathcal{H}$.
It has the complex structure $i_A$ and inner product $\innerProd[\mathcal{H}]{\cdot}{\cdot}$ as given in \eqref{eq:Fermionic.ComplexStructure} and \eqref{eq:Fermionic.InnerProduct}, respectively. 
The complex inner product $\innerProd[\mathcal{H}]{f}{g}$ of two elements $f, g \in \mathcal{H}$ has as integral kernel the two-point function
\begin{align}
\label{eq:Fermionic.Majorana.TwoPointFunction}
    w(x, y)
  = \bra{\Omega} \psi(x) \psi^\transpose(y) \ket{\Omega}
  &= \frac{1}{2}
    \begin{pmatrix}
      1 & 0 \\
      0 & 1
    \end{pmatrix}
    \updelta(x - y)
    - \frac{\i}{2} i_A(x, y)
  \eqend{.}
\end{align}

For the computation of the right wedge references, we use the Bisognano--Wichmann relation \eqref{eq:RightWedge.ModularOperatorLorentzBoost} on the Fock space, $\hat{U}(s) = \Gamma(\Delta)^{-\i s / (2 \pi)}$, where $\hat{U}(s)$ is the unitary representation of the boost acting on the Majorana fields as
\begin{align}
    \hat{U}(s) \psi(t, x) \hat{U}(s)^*
  &= \begin{pmatrix}
      \cosh\frac{s}{2} & - \sinh\frac{s}{2} \\
      - \sinh\frac{s}{2} & \cosh\frac{s}{2}
    \end{pmatrix}
    \psi\Bigl( \Lambda(s)(t, x) \Bigr)
  \eqend{,}
\end{align}
and $\Gamma$ denotes second quantization, so that the action of the modular Hamiltonian on the field operators becomes
\begin{subequations}
\eqseqlabel{eq:Fermionic.RightWedge.ModularAction.AtTimeZero.Computation}
\begin{align}
  \picklhs{
    - \i \log \Gamma(\Delta) \psi(x) \ket{\Omega}
  }
  &= 2 \pi \left. \deriv{}{s} \hat{U}(s) \psi(t, x) \hat{U}(s)^* \ket{\Omega} \right\rvert_{s = 0, t = 0}
\nexteq
  \skiplhs
  &\pickindent{=} 2 \pi
    \left.
      \left(
        \begin{pmatrix}
          0 & -\frac{1}{2} \\
          -\frac{1}{2} & 0
        \end{pmatrix}
        + x \one \partial_t
        + t \one \partial_x
      \right)
      \psi(t, x)
      \ket{\Omega}
    \right\rvert_{t = 0}
\nexteq
  \skiplhs
  &\pickindent{=} \pi
    \begin{pmatrix}
      0 &
      - 2 m x
      - 1
      - 2 x \partial_x
      \\
      - \left( -2 m x
      + 1
      + 2 x \partial_x \right) &
      0
    \end{pmatrix}
    \psi(x)
    \ket{\Omega}
  \eqend{.}
\end{align}
\end{subequations}
To find the expression for the corresponding one-particle modular Hamiltonian \eqref{eq:Fermionic.RightWedge.ModularGeneratorKernel.AtTimeZero} and its blocks \eqref{eq:Fermionic.RightWedge.ModularGeneratorKernel.Block}, we integrate against $\Reals^2$-valued test functions $f = (f_+, f_-), g = (g_+, g_-)$ in $x$ and let the derivatives act on the test functions using integration by parts.
For a compact notation of the operator's block form, we use the Pauli matrices $\sigma^1$ and $\sigma^2$, which have the properties $(\sigma^1)^\transpose = \sigma^1$ and $(\i \sigma^2)^\transpose = -\i \sigma^2$.
We compute
\begin{subequations}
\eqseqlabel{eq:Fermionic.RightWedge.ModularHamiltonian}
\begin{align}
  \picklhs{
    \innerProd{f}{- i_A \log \Delta g}
  }
  &= \bra{\Omega}\psi(f) \bigl( - \i \log \Gamma(\Delta) \psi \bigr)(g) \ket{\Omega}
\nexteq
  \skiplhs
  &\pickindent{=} \pi \sum_{i, j, k \in \{+, -\}}
    \int_{\Reals^2}
      f_i(x)
      g_j(y)
      \bra{\Omega}
      \psi_i(x)
  \eqbreakr\times
      \Bigl(
        - 2 m y (\i \sigma_2)_{j k}
        - (\sigma_1)_{j k}
        - 2 (\sigma_1)_{j k} y \partial_y
      \Bigr)
      \psi_k(y)
      \ket{\Omega}
    \id{x} \id{y}
\nexteq
  \skiplhs
  &\pickindent{=} \pi \sum_{i, j, k \in \{+, -\}}
    \int_{\Reals^2}
      f_i(x)
      w_{i k}(x, y)
  \eqbreakr\times
      \Bigl(
        2 m y (\i \sigma_2)_{k j}
        + (\sigma_1)_{k j}
        + 2 (\sigma_1)_{k j} y \partial_y
      \Bigr)
      g_j(y)
    \id{x} \id{y}
  \eqend{.}
\end{align}
\end{subequations}
So the one-particle modular Hamiltonian
\begin{align}
\label{eq:Fermionic.RightWedge.ModularHamiltonian.OneParticle}
    - i_A \log \Delta
  &= \begin{pmatrix}
      0 & M_- \\
      - M_+ & 0
    \end{pmatrix}
\end{align}
acts on vectors $h_+ \oplus h_- = h \in \mathcal{H}$ as
\begin{align}
\label{eq:Fermionic.RightWedge.ModularHamiltonian.OneParticleComponents}
    (M_{\pm} h_{\pm})(x)
  &= 2 \pi m x h_{\pm}(x)
    \mp \pi h_{\pm}(x)
    \mp 2 \pi x \partial_x h_{\pm}(x)
  \eqend{.}
\end{align}
Integration by parts then yields the operator kernels $M_{\pm}(x, y)$ as given in \eqref{eq:Fermionic.RightWedge.ModularGeneratorKernel.AtTimeZero}.

\section{Derivation of the skew-symmetric generator on the cylinder spacetime}
\label{appx:SkewSymmetricGenerator.Cylinder.Massive}

We make analogous calculations to the previous appendix for the cylinder spacetime (with circumference $l$). 
In the standard formalism, the field equations on the time-0 Cauchy slice are solved with the quasi-periodicity condition $\psi( x + l ) = \e^{\i \xi \pi} \psi( x )$, where $\xi \in \{ 0, 1 \}$ (for periodic and antiperiodic conditions, respectively).
We remark that $\xi$ could take real values in $[0, 2)$ more generally, but we only considered the two edge cases of periodic and antiperiodic boundaries.
To represent the field equation and the quasi-periodic functions in momentum space, we have to take the modified Fourier series
\begin{subequations}
\label{eq:FourierSeries.Generalized}
\begin{align}
    f(x)
  &= \frac{1}{l}
    \sum_{k = - \infty}^{\infty}
      \hat{f}( k )
      \e^{2 \pi \i \left( k + \frac{\xi}{2} \right) \frac{x}{l}}
  \eqend{,}
\nexteq
    \hat{f}( k )
  &\coloneq \int_{-\frac{l}{2}}^{\frac{l}{2}}
      f(x)
      \e^{- 2 \pi \i \left( k + \frac{\xi}{2} \right) \frac{x}{l}}
    \id{x}
  \eqend{.}
\end{align}
\end{subequations}
The differential operator of the field equations takes the same formal momentum space expression as in the case of Minkowski spacetime up to the different normalisation factor in the Fourier series. 
This leads to the series expansion of $S$ and $S_0$ as given in \eqref{eq:FermionicCylinder.SkewSymmetricGenerator.Series} and \eqref{eq:FermionicCylinder.SkewSymmetricGenerator.Series.Massless}, respectively, which we use for the Hilbert space of real-valued, square-integrable functions over the domain $[ -l/2, l/2 )$.
In the following, we show the computation of these Fourier series for the distributional operator kernels. 

First, let us consider a related, but simpler Fourier series that arises for a massive, scalar boson ($m > 0$, $\mu \coloneq \frac{m l}{2 \pi}$) on a cylinder spacetime with boundary conditions $\xi \in \{ 0, 1 \}$ so that $\cos( \xi \pi ) \in \{ 1, -1 \}$. 
The fundamental solution of the differential equation $A_{\mathrm{b}} \psi = ( - \partial_x^2 + m^2 ) \psi = 0$ (related to the Klein-Gordon equation) on the cylinder spacetime is the distribution
\begin{align}
\label{eq:BosonicCylinder.HelmholtzInverse}
    A^{-1}_{\mathrm{b}}(x, y)
  &= \frac{
      \sinh\bigl( m l - m \lvert x - y \rvert \bigr)
      + \e^{\i \xi \pi \sgn(x - y)} \sinh\bigl( m \lvert x - y \rvert \bigr)
    }{
      2 m \bigl( \cosh(m l) - \cos(\xi \pi) \bigr)
    }
  \eqend{,}
\end{align}
where the coordinate positions take values $x, y \in [ -l/2, l/2 )$.  
This inverse Helmholtz operator has the Fourier series 
\begin{align}
\label{eq:BosonicCylinder.HelmholtzInverse.Series}
    A^{-1}_{\mathrm{b}}(x, y)
  &= \frac{l}{( 2 \pi )^2}
    \lim_{\varepsilon \to 0^+} \sum_{k = -\infty}^{\infty}
      \frac{1}{\left( k + \frac{\xi}{2} \right)^2 + \mu^2}
      \e^{2 \pi \i \left( k + \frac{\xi}{2} \right) \frac{x - y}{l} - \varepsilon \abs{k}}
  \eqend{.}
\end{align}
Note that the $x$-derivative of this series
\begin{align}
\label{eq:BosonicCylinder.HelmholtzInverse.PositionDeriv}
    \pderiv{}{x} A^{-1}_{\mathrm{b}}(x, y)
  &= \frac{\i}{2 \pi}
    \lim_{\varepsilon \to 0^+} \sum_{k = -\infty}^{\infty}
      \frac{k + \frac{\xi}{2}}{\left( k + \frac{\xi}{2} \right)^2 + \mu^2}
      \e^{2 \pi \i \left( k + \frac{\xi}{2} \right) \frac{x - y}{l} - \varepsilon \abs{k}}
  \eqend{.}
\end{align}
is (up to a constant factor) identical to the mass-derivative of the skew-symmetric kernel $S( x, y )$ with Fourier series~\eqref{eq:FermionicCylinder.SkewSymmetricGenerator.Series}, 
\begin{subequations}
\eqseqlabel{eq:FermionicCylinder.SkewSymmetricGenerator.Series.MassDeriv}
\begin{align}
  \picklhs{
    \pderiv{}{m} S(x, y)
  }
  &= \pderiv{\mu}{m} \pderiv{}{\mu} S(x, y)
\nexteq
  \skiplhs
  &= \frac{l}{2 \pi} \frac{2 \i}{l}
    \lim_{\varepsilon \to 0^+} \sum_{k = -\infty}^{\infty}
      \frac{k + \frac{\xi}{2}}{\left( k + \frac{\xi}{2} \right)^2 + \mu^2}
      \e^{2 \pi \i \left( k + \frac{\xi}{2} \right) \frac{x - y}{l} - \varepsilon \abs{k}}
  \eqend{.}
\end{align}
\end{subequations}
By comparing with the derivative of \eqref{eq:BosonicCylinder.HelmholtzInverse}, we find the mass-derivative to be 
\begin{subequations}
\eqseqlabel{eq:FermionicCylinder.SkewSymmetricGenerator.MassiveIntegralKernel}
\begin{align}
	\picklhs{
		\pderiv{}{m} S( x, y )
	}
	&= 2 \pderiv{}{x} A^{-1}_{\mathrm{b}}( x, y )
\nexteq
	\skiplhs
	&= - \sgn( x - y )
		\frac{
			\cosh\bigl( m l - m \lvert x - y \rvert \bigr)
			- \cos( \xi \pi ) \cosh\bigl( m \lvert x - y \rvert \bigr)
		}{
			\cosh( m l ) - \cos( \xi \pi )
		}
	\eqend{,}
\end{align}
\end{subequations}
so that the kernel of $S$ for a generic mass parameter becomes 
\begin{align}
\label{eq:FermionicCylinder.SkewSymmetricGenerator.MassiveIntegral}
		S( x, y )
	&= S_{0}( x, y )
		+ \int_{0}^{m}
		\left. \pderiv{}{m} S( x, y ) \right\rvert_{m = \tilde{m}}
		\id{\tilde{m}}
	\eqend{,}
\end{align}
with the special cases as given in~\eqref{eq:FermionicCylinder.SkewSymmetricGenerator.MassiveIntegral.SpecialCases}.
The integrand is a smooth function in $\tilde{m}$, monotonically decreasing and vanishing at infinity from the initial value 
\begin{align}
\label{eq:FermionicCylinder.SkewsymmetricGenerator.MassiveIntegral.MasslessLimit}
    \lim_{\tilde{m} \to 0^+}
      \left. \pderiv{}{m} S( x, y ) \right\rvert_{m = \tilde{m}}
  &= \begin{dcases}
      - \sgn( x - y ) \left( 1 - \frac{2 \abs{x - y}}{l} \right)
      & \text{~if~} \xi = 0
      \eqend{,}
    \\
      - \sgn( x - y )
      & \text{~if~} \xi = 1
      \eqend{.}
    \end{dcases}
\end{align}
The massive correction integral in \eqref{eq:FermionicCylinder.SkewSymmetricGenerator.MassiveIntegral}
only contributes for $m > 0$, otherwise $S( x, y )$ reduces to $S_0( x, y )$ as given in~\eqref{eq:FermionicCylinder.SkewSymmetricGenerator.Series.Massless}.

\section{Analytic references for massless fields on double cones in the cylinder spacetime}
\label{appx:Cylinder.MasslessModularHamiltonian}

The massless modular Hamiltonian has been calculated for Dirac fermions on a general multi-component subspace region in 2-dimensional Minkowski spacetime~\cite{CasiniHuerta:2009}.
This work was later generalised to the 2-dimensional cylinder spacetime~\cite{KlichVamanWong:2017}.
The modular Hamiltonian as computed in these publications is restricted to the subspace region and it is closely related to the modular generator $-\log \varDelta$ on the subspace, see~\cite[Sec.~2]{CadamuroFroebMinz:2025}.
Since the complement of the region $\mathcal{B}$ is again a union of intervals, these results extend to the full Hilbert space.
We use them as analytic references in the massless cases.
Even though the results in \cite{KlichVamanWong:2017} are for Dirac fermions, these can be understood as complex-linear combinations of Majorana fermions, allowing us to find the corresponding expressions for Majorana fermions used in the present paper.

For a subspace region over a set of $n$ non-overlapping intervals $\mathcal{B} = \bigcup_{i = 1}^{n}[a_i, b_i]$ on the circle (as Cauchy surface of the cylinder spacetime), the modular Hamiltonian is written in terms of a function $z : [0, l) \to \Complexes$~\cite{KlichVamanWong:2017},
\begin{align}
\label{eq:FermionicCylinder.MulticomponentFactorGenerator}
    \e^{z(x)}
  &= \lim_{\varepsilon \to 0^+} (-1)^{n - 1} 
    \prod_{j = 1}^{n}
      \frac{
        \exp\Bigl( 2 \pi \i \frac{x + \i \varepsilon}{l} \Bigr)
        - \exp\Bigl( 2 \pi \i \frac{a_j}{l} \Bigr)
      }{
        \exp\Bigl( 2 \pi \i \frac{b_j}{l} \Bigr)
        - \exp\Bigl( 2 \pi \i \frac{x + \i \varepsilon}{l} \Bigr)
      }
  \eqend{.}
\end{align}
The equation $z(x) = z(y)$ has $n$ distinct solutions $y = v_k(x)$, where $k \in \{ 0, \dots,n - 1 \}$ and where $v_0(x) = x$ is the trivial solution.
For a shorter notation, define the interval widths $w_i \coloneq b_i - a_i$, the interval mid-points $\frac{s_i}{2} \coloneq \frac{a_i + b_i}{2}$, and the total length of the subspace region $w \coloneq \sum_{i = 1}^{n} w_i$. 
For any periodicity parameter $\xi \in \{0, 1\}$, there are two contributions to the modular Hamiltonian, and a ``zero mode'' term $H_{0}$ only for $\xi = 0$ (denoted with a Kronecker delta)~\cite{KlichVamanWong:2017},
\begin{subequations}
\eqseqlabel{eq:FermionicCylinder.MultiComponentRegion.ModularHamiltonian}
\begin{align}
    \i H(x, y)
  &= \i H_{\mathrm{loc}}(x, y)
    + \i H_{\mathrm{biloc}}(x, y)
    + \i \updelta_{\xi 0} H_{0}(x, y)
  \eqend{,}
\nexteq
\label{eq:FermionicCylinder.MultiComponentRegion.ModularHamiltonian.Local}
    \i H_{\mathrm{loc}}(x, y)
  &\pickindent{=}
    \pi \left( z'(x)^{-1} + z'(y)^{-1} \right)
    \updelta'(x - y)
  \eqend{,}
\nexteq
\label{eq:FermionicCylinder.MultiComponentRegion.ModularHamiltonian.Bilocal}
    \i H_{\mathrm{biloc}}(x, y)
  &= \frac{2 \pi^2}{l}
    \sum_{k = 1}^{n - 1}
    \e^{\i \xi \pi \frac{x - y}{l}}
    \left( \i \updelta_{\xi 0} - \cot\left( \pi \frac{x - y}{l} \right) \right)
    z'(y)^{-1}
    \updelta\bigl( v_k(x) - y \bigr)
  \eqend{.}
\end{align}
\end{subequations}
The ``local'' term $-\i H_{\mathrm{loc}}$ is supported along the diagonal $y = x$, the ``bilocal'' terms $-\i H_{\mathrm{biloc}}$ correspond to the correlations along the curves $y = v_k( x )$ for $k \in \{ 1, \dots,n - 1\}$.

Note that the ``zero mode'' contribution considered in \cite{KlichVamanWong:2017} does not correspond to the state that we implemented.
Instead, we considered the vacuum state corresponding to the limit $m \to 0$, in which the kernel of $H_0$ is \cite[Cor.~2]{CadamuroFroebGuillem:2024}
\begin{align}
\label{eq:FermionicCylinder.MultiComponentRegion.ModularHamiltonian.Zeromode}
    \i H_0(x, y)
  &= \i \pi
    \frac{
      \cos\bigl( \pi \frac{x - y}{l} \bigr)
      - \cos\bigl( \pi \frac{w_1}{l} \bigr)
    }{
      \sin\bigl( \pi \frac{w_1}{l} \bigr)
    }
    \updelta(x + y - s_1)
\end{align}
symmetric in its variables $x$ and $y$ and supported along the antidiagonal $y = -x + s_1$, for a single interval $[a_1, b_1]$.

Different choices of boundary conditions correspond to different sectors, periodic boundaries are associated to the Ramond sector, while antiperiodc boundaries belong to the vacuum or Neveu-Schwarz sector~\cite[Sec.~3.6]{Fuchs:1992}.
Note that in some works like~\cite{RehrenTedesco:2013}, the fields are rescaled by a phase factor in comparison to our notation, such that in that context, the Neveu-Schwarz sector is understood to belong to periodic boundaries.

The modular Hamiltonian $H$ is Hermitian, hence the real part of the kernel $\i H(x, y)$ is skew-symmetric in $x$ and $y$ and corresponds to the skew-symmetric part of the expression for Majorana fermions, while the imaginary part is symmetric in $x$ and $y$ and is given by the symmetric part of the Majorana expression,
\begin{align}
    \i H(x, y)
  &= M_{-, \mathrm{skew}}(x, y)
    + \i M_{-, \mathrm{sym}}(x, y)
\end{align}
or inverted to 
\begin{align}
    M_-(x, y)
  &= \Re\Bigl( \i H(x, y) \Bigr)
    + \Im\Bigl( \i H(x, y) \Bigr)
  \eqend{.}
\end{align}
The profile functions $z'( x )^{-1}$ for different subspace setups --- for an individual double cone given in \eqref{eq:FermionicCylinder.OneDoubleCone.MulticomponentFactor} and for two double cones given in \eqref{eq:FermionicCylinder.TwoDoubleCones.MulticomponentFactor} --- follows from \eqref{eq:FermionicCylinder.MulticomponentFactorGenerator} when inserting the corresponding interval bounds $a_i$ and $b_i$, solving for $z(x)$, taking its first derivative and computing the reciprocal.

\printbibliography

\end{document}